\documentclass[preprint,sort&compress]{elsarticle}
\usepackage{amssymb,amsthm,amsmath,commath,bbm,setspace,color,array,booktabs,multirow,url,float,wasysym,algorithm,algpseudocode,mathtools,bm}
\usepackage{graphicx}
\usepackage{diagbox} % for diagonal line in table, alternative slashbox with syntanx \slashbox
\usepackage[table]{xcolor}
\usepackage[FIGTOPCAP,nooneline]{subfigure}
\usepackage[normalem]{ulem}
\usepackage{epstopdf}
\usepackage{url}
 
\allowdisplaybreaks

\makeatletter
\def\ps@pprintTitle{%
 \let\@oddhead\@empty
 \let\@evenhead\@empty
 \let\@evenfoot\@oddfoot}
\makeatother

\usepackage[colorlinks,
    linkcolor={red!50!black},
    citecolor={red!50!black},
    urlcolor={blue!80!black}]{hyperref}
    
\usepackage{tikz,pgf,pgffor,pgfplots}
\tikzset{>=latex}
\usetikzlibrary{calc,patterns,decorations,intersections,decorations.pathmorphing,fit,backgrounds,arrows,shapes,snakes,automata,petri,positioning,graphs,graphs.standard,decorations.pathreplacing}
\usepgfmodule{shapes,plot}

\numberwithin{equation}{section} %\numberwithin{figure}{section}
\numberwithin{table}{section}

\usepackage[colorinlistoftodos]{todonotes}
\newcommand{\x}{\textnormal{\textbf{x}}}
\newcommand{\y}{\textnormal{\textbf{y}}}
\newcommand{\z}{\textnormal{\textbf{z}}}
\newcommand{\bb}{\textnormal{\textbf{b}}}

\newcommand{\p}{\textnormal{\textbf{p}}}
\newcommand{\ub}{\textnormal{\textbf{U}}}

\newcolumntype{P}[1]{>{\centering\arraybackslash}p{#1}}
\begin{document}
\begin{frontmatter}
 \title{Finite difference physics--informed neural networks enable improved solution accuracy of the Navier--Stokes equations}
\author[SRM]{Nityananda Roy\corref{cor}}
\cortext[cor]{Corresponding author}
\ead{nityananda.r@srmap.edu.in}
\author[MAG]{ Robert Dürr}
\author[ELR]{Andreas Bück}
\author[IIT]{S. Sundar}

\address[SRM]{Department of Mathematics, SRM University-AP, Amaravati, Andhra Pradesh 522502, India}
\address[MAG]{Engineering Mathematics, Magdeburg--Stendal University of Applied Sciences, Breitscheidstraße 2,
39114 Magdeburg, Germany}
\address[ELR]{Institute of Particle Technology, Friedrich--Alexander--University Erlangen--Nuremberg, Cauerstrasse 4, 91058 Erlangen, Germany}
\address[IIT]{Department of Mathematics, Indian Institute of Technology Madras, Chennai 600036, India}

\begin{abstract}
Generating an accurate solution of the Navier--Stokes equations using physics--informed neural networks (PINNs) for higher Reynolds numbers in the corners of a lid--driven cavity problem is challenging.  In this paper, we improve the solution accuracy of the incompressible Navier--Stokes equations in the region near the walls significantly and generate accurate secondary vortices in the corners of the lid--driven cavity by solving the governing equations using finite difference--based PINNs (FD--PINNs) without employing the known solution.  We adopt the domain decomposition method (DDM) and combine it with the FD--PINNs to solve the lid--driven cavity problem for the Reynolds numbers Re = 400 and Re=1000. A comparison of the mean square error (MSE) between the presented and standard FD--PINNs using the reference solution is exhibited, showing the accuracy and effectiveness of the new approach.
\end{abstract}
\end{frontmatter}

\section{Introduction}

In the past several years, interest in simulation of fluid flow in a lid--driven cavity has grown due to its flow structure being significant across various fields, such as engineering to bio--medical \cite{migeon2003} and mixing process \cite{huang2019mixing} etc. Due to the simple geometrical structure of the problem, it provides benchmark data for validation and other comparison.
\begin{figure}[htbp!]
\centering
\begin{tikzpicture}[scale=1,transform shape] %incline plane
triangle coordinates
\coordinate (A) at (0,5);
\coordinate [label=below:$\substack{0}$] (B) at (-0.05,-0.05);
\coordinate [label=below:$\substack{1}$] (B) at (5.1,-0.05);
\coordinate [label=below:$\substack{1}$] (B) at (-0.05,5.4);
\coordinate [label=below:$\substack{\text{x}\vspace{0.2 cm}}$] (B) at (2.5,-0.1);
\coordinate [label=right:$\substack{\vspace{0.2 cm}}$] (C) at (5.1,2.5);
\coordinate [label=above:$\substack{}$] (D) at (2.5,5.1);
\coordinate [label=left:$\substack{\text{y}\vspace{0.2 cm}}$] (E) at (-0.1,2.5);
%\coordinate [label=left:$\substack{X}$] (F) at (-0.2,2.5);
%\coordinate [label=above:$\nu\,\text{=}\,10^{-6}$ m$^2$\slash s] (E) at (2.6,3.2);
%\coordinate [label=below:at $20^{\circ}\text{C}$] (E) at (2.6,3.2);
%\coordinate [label=above:Lid driven cavity $\rightsquigarrow$ "Easier"] (C) at (2.5,-6.5);

\draw (A) -- ++(0,-5) -- ++(5,0) -- ++(0,5);
\draw[->,blue,line width=0.5mm] (-0.5,5) -- ++(6.,0);
\draw[<-,blue,line width=0.5mm] (1.17,3.5) arc (-10:270:-1.5);
\draw[->,blue,line width=0.5mm] (1.2,0.7) arc (-10:300:0.5);
\draw[->,blue,line width=0.5mm] (4.7,1) arc (30:350:0.5);
\draw[clip] (0,0) rectangle (5,5);
%\foreach \i in {1,...,500}
%  \fill (rnd*5cm, rnd*5cm) circle (1pt);
% \draw[fill=blue!50!green] (2.3,0) -- (2.8,1) -- (3,0.9) -- (2.55,0) -- cycle;
% \draw[blue!50!green,line width=0.5mm] (2.3,0) -- (2.8,1) -- (3,0.9) -- (2.55,0);
\end{tikzpicture}
\caption{Geometry of a 2D square lid--driven cavity.}
\label{fig:cavity}
\end{figure}
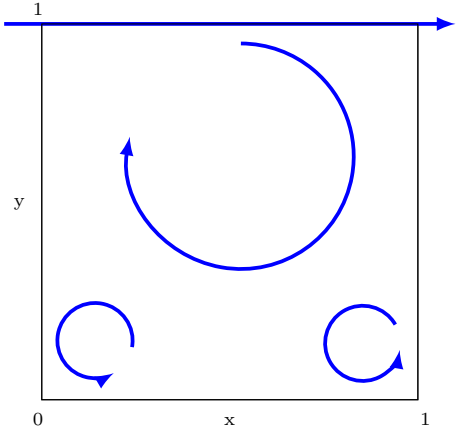
However, flow physics inside a lid--driven cavity is not straightforward. The flow within the driven cavity demonstrates several interesting physical characteristics, such as boundary layers, eddies of different sizes, and instability, which also appear in processing industries \cite{peng2003}. In addition, lid--driven cavity flow demonstrates almost all physical phenomena that can occur in incompressible flow, such as secondary vortex, complex flow pattern, chaotic motion, turbulence, etc. Depending on the Reynolds numbers, the lid--driven cavity flow holds counter--rotating secondary vortices in the corners of the bottom of the cavity; see Figure \ref{fig:cavity}. The formation of those is usually related to the complex interaction of the fluid flow and geometry of the cavity. It is also an outcome of the interaction between the primary vortex and the boundary conditions of the fluid flow. In particular, the change in the flow direction in the corner of the cavity is the reason behind the creation of this vortex. Furthermore, the size of the secondary vortex depends on the Reynolds number, the aspect ratio of the cavity, and other parameters. Finding an accurate representation of the secondary vortex in simulations of lid--driven cavity flow has been a research interest for several decades. Previous attempts used finite difference methods \cite{ebrahimijahan,zogheib2011,tian2011efficient}, finite element methods 
 \cite{hachem2010,coupez2013solution,lim2017nonconforming}, finite volume methods \cite{sahin2003novel,sousa2016lid,magalhaes2013}, and lattice Boltzmann methods \cite{sun2020high,perumal2011multiplicity,wu2004simulation}, that achieved satisfactory representation of the secondary vortices.
 
\par Increasing availability of computational resources enabled deep learning--based methods to solve various types of partial differential equations (PDEs) as they can handle high dimensional and non--linear problems. Usually, two types of deep learning algorithms are applied to obtain approximate solutions to non--linear PDEs: data--driven and data--free. One significant advantage of data--free methods over data--driven ones is overcoming the issue of an insufficient amount of data to train the model. Recently, PINNs \cite{raissi2019physics} has been developed to solve forward and inverse PDE problems. The basic idea of this approach is minimizing a loss function by constraining the governing equation along with the initial and boundary conditions. In this method, no data is required to train the governing equation component of the loss function; data is necessary only for the initial and boundary conditions, which are usually given with the governing equations. Automatic differentiation (AD) \cite{baydin2018automatic} is used to evaluate the differentiation terms.
 \par PINNs has been applied for the solution of different flow problems, some of which are discussed herein: solution of the incompressible Navier–Stokes equations using PINNs by considering two forms of the governing equations, velocity–pressure, and velocity–vorticity, is carried out in \cite{jin2021nsfnets}. Mao et al. \cite{mao2020physics} employed PINNs to solve the one--and two--dimensional Euler equation to analyze aerodynamic flows. Eivazi et al. \cite{eivazi2022physics} solved the Reynolds--averaged Navier--Stokes equations for incompressible turbulent flows using PINNs without considering any specific model for turbulence. Cai et al. \cite{cai2021heat} employed physics--informed neural networks to solve a heat transfer flow problem with a moving interface. A detailed review of PINNs for several multiphysics and multiscale flow problems in two and three--dimensional cases has been presented in \cite{cai2021physics}. PINNs is suitable for any boundary condition, but sometimes trained networks fail to satisfy the target boundary condition, especially for complex models. The accuracy of automatic differentiation is coupled with the accuracy of the neural network. Automatic differentiation is dependent on the explicit functional expression of the output of the neural network with respect to input, i.e., the derivatives are precisely estimated only when the neural networks accurately approximate the ground truth. Besides, computational cost increases significantly while calculating derivatives using automatic differentiation when a large number of layers and neurons are considered, which is mostly a necessity to generate a convergent solution. 

\par Due to the aforementioned reasons, PINNs fail to render adequate accuracy compared to classical numerical methods for high Reynolds numbers and are also unable to generate secondary vortices in the bottom of the cavity \cite{jiang2023}, while solving the Navier–Stokes equations in a lid–driven cavity. Therefore, several researchers improved standard PINNs to achieve satisfactory results for high Reynolds number flows. One recent development is presented in \cite{chiu2022can}, where a coupled automatic–numerical differentiation is used in PINNs (can–PINNs) instead of just automatic differentiation to combine the advantages of both automatic and numerical differentiation. It is shown that can--PINNs can provide better results than the AD--PINNs for the case of Re=400 without using the known solution. However, with the utilization of the known solution, the mean square error has been discussed for the Reynolds number Re=1000. Jiang et al. \cite{jiang2023} applied finite difference--based PINNs by utilizing the known solution generated from the classical method to solve Navier--Stokes equations for high Reynolds numbers (Re=1000 and Re=5000) in a lid–driven cavity problem. Here, a detailed comparison of accuracy of FD--PINNs compared to AD--PINNs was presented. They used known solutions generated from classical numerical methods as sample points and used it as an additional component of the loss function to improve the accuracy of the solution. With the inclusion of sample points, FD–PINNs exhibit better results. Without the sample points, FD–PINN failed to generate accurate secondary vortices in the lower corners of the cavity for Re=1000. To improve accuracy, Xiao et al. \cite{xiao2024least} replaced the traditional finite difference with the least squares--based finite difference to calculate the derivative of the differential terms. In other words, the primary focus was to improve solution accuracy by improving the accuracy of the derivative of the differential terms. The traditional finite difference method has been considered near the boundary region as least--squares finite difference methods face challenges at these locations. In contrast, the least squares--based finite difference has been used inside the domain. Although this process provides quite accurate results, it is computationally complex to implement. In addition, the accuracy of the derivative is sensitive to the weighting function and supporting points.

\par Recently, domain decomposition methods (DDM) combined with PINNs has been developed to obtain accurate solutions of non--linear PDEs. The basic idea is to divide the entire spatial domain into small sub--domains and solve PDEs on each of the sub--domains, i.e., the solution of the entire problem converts into sub--domain problems. The solution of each of the sub--domains can be proceeded independently and in parallel. Combination of DDM with PINNs has been investigated in several publications, e.g., \cite{heinlein2021, li2019d3m,wu2022improved,jagtap2020extended,shukla2021parallel,gu2024physics}. A recent development is a combination of AD--PINNs with DDM for solving unsteady Navier--Stokes equations in a 2D lid--driven cavity problem for cases Re=100 and Re=400 \cite{gu2024physics}. The study considered stream function--pressure formulation of the governing equations. Two types of learning strategies were applied to generate solutions: (i) learning from scratch and (ii) transfer learning. Transfer learning utilized trained neural networks for the low Reynolds number (Re=100) to solve Navier--Stokes equations for the higher Reynolds number (Re=400). This approach lead to better accuracy than learning from scratch. Based on the streamfunction--pressure formulation, transfer learning based  AD--PINNs coupled with DDM provide accurate results for the Reynolds number Re=100, but accuracy is reduced for the Reynolds number Re=400.
\par In the discussion above, it has been noted that some methods can generate secondary vortices by utilizing the known solution generated from the classical method or at the cost of high computational complexity, and some fail. The primary objective of this paper is to develop a straightforward technique based on a combination of FD--PINNs and DDM, without employing the known solutions at sample points to improve the solution accuracy of the Navier--Stokes equations. Specifically, the case Re=1000 is considered, focusing on near the walls and in the corners of the lid--driven cavity where secondary vortices appear. Instead of solving in each of the sub--domains in parallel, first, we solved the velocity--pressure form of steady Navier--Stokes equations in the entire domain by training neural networks using FD--PINNs and used this trained weight and bias as the initial conditions to generate solutions for the small sub--domains. Therefore, no interphase conditions are required in this process to generate solutions, which reduces computation and training complexity significantly. The boundary conditions for each sub--domain need to be obtained from the solution of the entire domain. In this process, the solution of each sub--domain is completely independent of neighboring sub--domains. By comparison with the reference solution we show the accuracy of the present technique. 

This paper is organized as follows: In Section \ref{sec:1}, we discuss the governing Navier--Stokes equations. In Section \ref{sec:2}, we explore finite difference--based PINNs and discuss the proposed technique to solve the Navier--Stokes equations for lid--driven cavity problems along with a detailed algorithm. Numerical results are discussed in Section \ref{sec:3}. Conclusions and outlook on future works are offered in Section \ref{sec:4}.

\newpage

\section{Navier--Stokes equations for lid--driven cavity flow}\label{sec:1}

The steady incompressible Navier–Stokes equations are given by \cite{bruneau1990}

\begin{align}
 \nabla \cdot \ub&=0 \label{eq:1.1}\\
 \ub \cdot \nabla \ub&=-\nabla p+\frac{1}{{Re}}\Delta \ub \label{eq:1.2}
\end{align}
%    \ub \cdot \nabla \ub=-\nabla p+\frac{1}{{Re}}\Delta \ub, \quad \text{where}\quad \nabla \cdot \ub=0.
%\end{equation}

The velocity and pressure of the fluid at the spatial position $\x =(x,y) \in \mathbb{R}^{2}$ are delineated by $\ub(\x)= \big(u(x,y),v(x,y)\big)$ and $p(x,y)$, respectively. The Reynolds number $Re$ is a dimensionless number determined by the characteristic velocity, characteristic length and viscosity of the fluid. It is observed that the left--hand side of the equation \eqref{eq:1.2} is a nonlinear term in  $\ub$. The incompressible condition is reflected as a constraint in the equation \eqref{eq:1.1}. There is no explicit governing equation for the pressure; therefore, pressure can be treated as a hidden state generated by using the incompressible condition. The geometry we have considered here is a square lid--driven cavity; see Figure \ref{fig:cavity}, in which the upper lid of the cavity is moving in the horizontal direction, and the remaining wall boundaries are rigid. Therefore, the dimensionless horizontal component of the fluid velocity at the top of the cavity can be defined as $u(x,y=1)=1$, while velocity on the other walls remains zero.

\section{Finite difference physics--informed neural networks (FD--PINNs)}\label{sec:2}
\begin{figure}[H]
    \centering
     \includegraphics[width=0.7\textwidth]{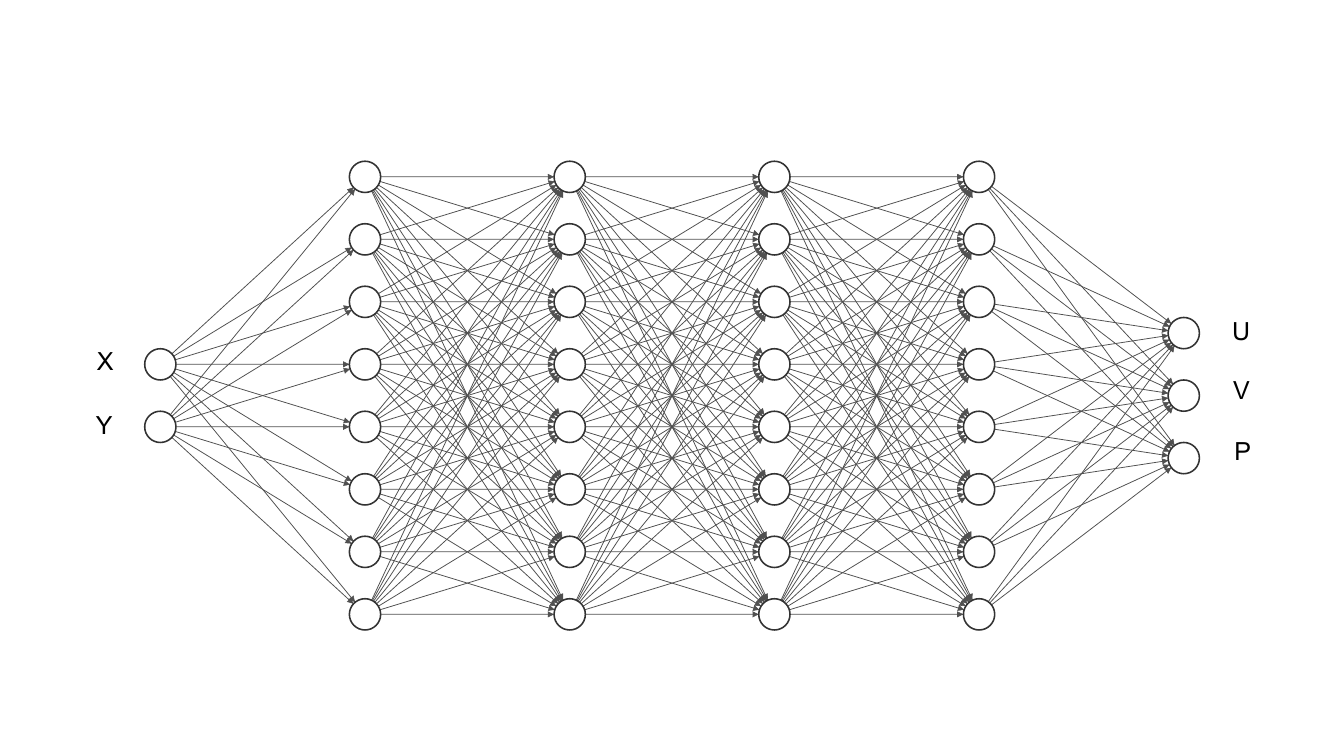}
    \caption{Neural networks with four hidden layers and eight neurons in each. }
    \label{fig:1}
\end{figure}

In this section, we explore finite difference--based PINNs in order to obtain the solution of the posed problem using the spatial coordinates as input. To obtain an accurate solution, fully connected neural networks with multiple hidden layers are considered, which are connected by a non--linear activation function. A hyperbolic $tanh$ function is considered for the nonlinear activation function to obtain a convergent solution. Other activation functions, such as $ReLU$ and $Swish$, are not considered in this study due to the occurrence of the vanishing gradient problem, whereas, $tanh$ activation effectively handles the vanishing gradient problem even for complex networks. For more details about the activation function, we refer to \cite{sharma2017,jagtap2020locally}. Generally, the differential terms in the PDE are evaluated using automatic differentiation of the output variable with respect to the input data \cite{baydin2018automatic}. For FD--PINNs, the standard finite difference method is applied to discretize the derivative terms. A pictorial representation of the neural--network, including all the layers, input, and output variables, is given in Figure \ref{fig:1}.

The input and output data are represented by $\z_{0} \in \mathbb{R}^{n_{0}}~ \text {and}~ \z_{K} \in \mathbb{R}^{n_{K}}$, respectively and the output data of the $l$th layer is denoted by $\z_{l}$, where $l = 1, 2, \cdots , K-1$. Two neighboring layers are linked by the following relation:

\begin{equation}\label{eq:3}
    \z_{l}=\tau_{l}(\omega_{l}^{\text{T}}\z_{l-1}+\bb_{l}), \quad \text{where} \quad l = 1, 2, \cdots , K-1.
\end{equation}
The weight matrix, bias and activation function are denoted by $\omega_{l}\in \mathbb{R}^{n_{l-1}\times n_{l}}$, $\bb_{l} \in \mathbb{R}^{n_{l}}$, and $\tau_{l}$, respectively, where $l$ is the index of each layer. The output of the neural networks $[\ub_{\mathcal{N}} ~~ p_{\mathcal{N}}]^{\text{T}}\approx \z_{K}(\x; \omega,\bb)=\omega_{K}^{\text{T}}\z_{K-1}+\bb_{K},$ can be utilized to approximate the solution of the equations (\ref{eq:1.1}--\ref{eq:1.2}), where $\omega=\{\omega_{l} \mid l=1,2,\cdots,K \}$ and $\bb=\{\bb_{l} \mid l=1,2,\cdots,K \}$ stand for the weight and bias of the entire network, respectively. The loss function of the minimization problem needs to be calculated at the interior points of the cavity, and in each iteration, the boundary conditions are added separately before applying the finite difference discretization of the differential terms. As \cite{jiang2023}, the boundary condition for the pressure is not applied as the pressure can be updated in each iteration of the training process. Therefore, no explicit loss function is required for the boundary condition. The total loss function can be defined by 
\begin{align}\label{eq:4}
    \mathcal{L}_{F_n} ({\Theta})&=\frac{1}{n_{x}\times n_{y}}\sum_{i=1}^{n_{x}}\sum_{j=1}^{n_{y}}\Big|\ub_{\mathcal{N}} \cdot \nabla\ub_{\mathcal{N}}+\nabla\p_{\mathcal{N}}-\frac{1}{{Re}}\Delta\ub_{\mathcal{N}}\Big|_{i,j}^{2} + \frac{1}{n_{x}\times n_{y}}\sum_{i=1}^{n_{x}}\sum_{j=1}^{n_{y}}\Big|\nabla\cdot \ub_{\mathcal{N}}\Big|_{i,j}^{2}.
\end{align}
where, the total number of interior points is $n_x\times n_y$. The optimal values of the networks' parameters $\Theta=\{\omega,\bb \}$ are computed via an appropriate optimization routine. Thereafter, the trained weight and bias are used to obtain the solution at any grid point of the equations (\ref{eq:1.1}--\ref{eq:1.2}).

A finite difference method has been applied to find the derivative term involved in the Navier--Stokes equations on the rectangular grid. The discretization of the differential terms using the finite difference method with second--order accuracy for the horizontal component of the velocity of the Navier--Stokes equations is defined as follows:

\begin{align*}
    \bigg[\frac{\partial u}{\partial x}\bigg]_{i,j}&=\frac{u(i+1,j)-u(i-1,j)}{2\Delta x}, \quad
    \bigg[\frac{\partial u}{\partial y}\bigg]_{i,j} =\frac{u(i,j+1)-u(i,j-1)}{2\Delta y},\\
     \quad 
     \bigg[\frac{\partial^{2} u}{\partial x^{2}}\bigg]_{i,j} &=\frac{u(i+1,j)-2u(i,j)+u(i-1,j)}{{\Delta x}^{2}},
     \quad
     \bigg[\frac{\partial^{2} u}{\partial y^{2}}\bigg]_{i,j}=\frac{u(i,j
     +1)-2u(i,j)+u(i,j-1)}{{\Delta y}^{2}},
\end{align*}

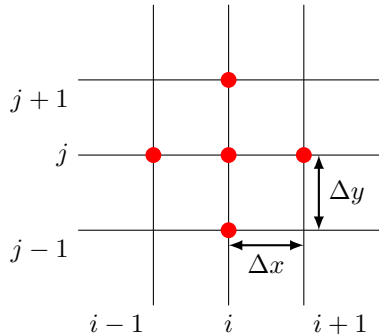
\begin{figure}[H]
\centering
\begin{tikzpicture}
    
    \foreach \x in {0,1,2,3} {
        \foreach \y in {0,1,2,3} {
            
            \ifnum\x>0
                \draw (\x,\y) -- (\x,\y+1); 
            \fi
           
            \ifnum\y>0
                \draw (\x,\y) -- (\x+1,\y); 
            \fi
        }
    }
   
  \draw[<->, thick] (2, 1-0.2) -- (3, 1-0.2) node[midway, below] {$\Delta{x}$};

    % Draw and label dy (vertical spacing)
    \draw[<->, thick] (3+0.2, 1) -- (3+0.2, 2) node[midway, right] {$\Delta{y}$};
    % Add labels on the edges
    \node foreach \j in {1} [below left]at (0,\j) {$j-1$};
    \node foreach \j in {2}  [left]at (0,\j) {$~j$};
    \node foreach \j in {3}  [below left] at (0,\j) {$j+1$};
    \node foreach \i in {1} [below left] at (\i,0) {$i-1$};
    \node foreach \i in {2} [below ] at (\i,0) {$i$};
    \node foreach \i in {3} [below right] at (\i,0) {$i+1$};
%    \node foreach \i in {1,2} [above] at (\i,3) {$20^\circ$};

    % Label cells with their indices
%    \node foreach \i in {1,2} [above right] at (\i,2) {$\i$};
%    \node foreach \lab [count=\n] in {4,3} [above right] at (\n,1) {$\lab$};

    % Add cyan points as specified
  \foreach \list[count=\j from 0] in {{},{},{3}} 
    \foreach \x in \list {
        \fill[red] (1,\j) circle (3pt);
    }

\foreach \list[count=\j from 0] in {{},{1},{2},{3}} 
    \foreach \x in \list {
        \fill[red] (2,\j) circle (3pt);
    }

\foreach \list[count=\j from 0] in {{},{},{3}} 
    \foreach \x in \list {
        \fill[red] (3,\j) circle (3pt);
    }
    
\end{tikzpicture}

 \caption{Framework for second--order central difference scheme}
    \label{fig:central}

\end{figure}

where $i,j$ refer to the indices of the grid point; see Figure \ref{fig:central}. The velocity component $u$ is the output data of the neural network, i.e., $u=u_{\mathcal{N}}$. In a similar manner, the discretized equation for the vertical velocity component is generated.  We used L--BFGS (\cite{liu1989limited}) algorithm to minimize the loss function. The key features of using the L--BFGS algorithm are its efficiency and faster convergence rate. Instead of calculating the first order derivative of the loss function, second--order derivatives derived from approximate Hessian matrix, makes it more effective and efficient than the usual gradient descent method. In addition, L--BFGS utilizes minimum memory allocation, which renders it suitable for higher dimensional problems and large--scale PINNs with a large number of inputs. 

\begin{figure}[H]
\centering
    \includegraphics[width=.65\textwidth]{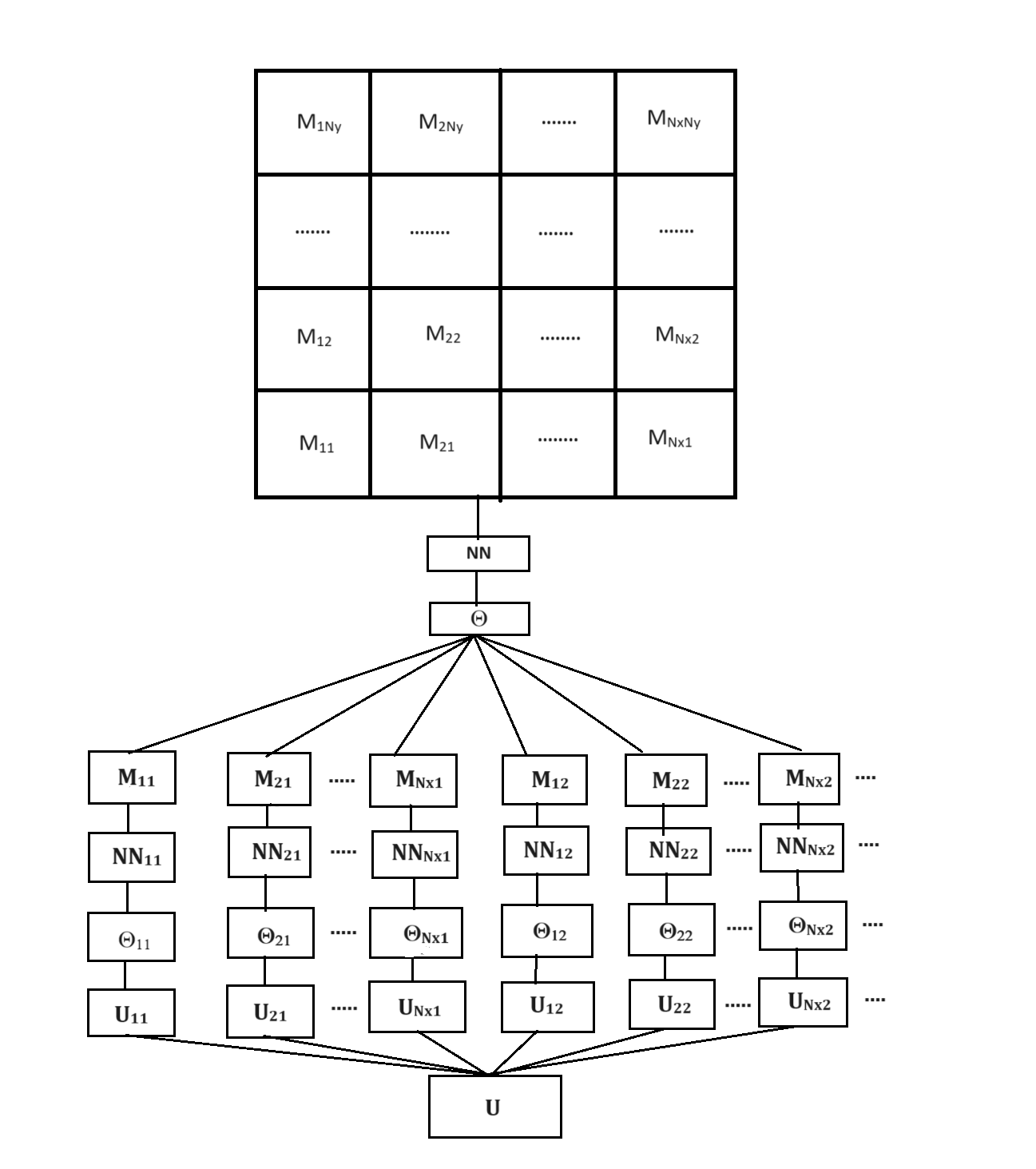} 
  \caption{Pictorial representation of the present technique}
  \label{fig:2}
\end{figure}

 In FD--PINNs, increasing the number of grid points leads to higher computational costs and raises other complexity, such as a decrease in solutions' accuracy \cite{jiang2023}. To avoid this issue, we adopt a DDM--based technique coupled with FD--PINNs but in a different manner. The basic idea relies on instead of solving the governing equations for very finer grid points using FD--PINNs, solving the equations in the entire domain for a sufficient grid points (can be determined through computational trials in the context of achieving the desired accuracy), which is followed by a process of dividing the whole domain into smaller sub--domains and train new neural networks in each of the sub--domains. The trained weight and bias of the neural networks for an entire domain are considered as the initial weight and bias in the training process of the new neural networks for the sub--domains; see Figure \ref{fig:2}. The boundary conditions for each sub--domain are obtained from the solution of the entire domain, i.e., no interphase conditions are required between neighboring sub--domains, and the training process for each sub--domain is completely independent. When the training process is finished for individual sub--domains, the sub--domain solutions are combined to obtain a solution for the entire lid--driven cavity, see algorithm \ref{alg:1}.

\begin{center}
\begin{minipage}{0.95\linewidth}
\begin{algorithm}[H]
    \begin{algorithmic}[1]
        \caption{Algorithm to solve steady Navier--Stokes equations\label{alg:1}} 
        \Require  Discretize the entire domain into rectangular grids and use these grid points as inputs. Set the initial conditions of $\omega$ using He initialization \cite{huang2022hompinns, he2015delving} and $\bb$ as zeros.
         \While {$iter~ <~ epoch$}
         \State Calculate $[\ub_{\mathcal{N}}~p_{\mathcal{N}}]^{T}\approx \z_{K} =\omega_{K}^{\text{T}}\z_{K-1}+\bb_{K}$ using $ \z_{l}$ where $l=1,\cdots,K-1$.
         \State Reshape the output variables to a matrix form and add the boundary of the velocity components.
         \State Compute derivatives of the differentiable terms involved in the PDEs using the finite difference methods. 
         \State  Calculate the total loss function using \eqref{eq:4} along with the incompressible condition.
         \State Using L--BFGS algorithm, minimize the loss function to obtain the values of $\omega=\{\omega_{l} \mid l=1,2,\cdots,K \}$ and $\bb=\{\bb_{l} \mid l=1,2,\cdots,K \}$.
         \State $iter=iter+1$
         \EndWhile
         \State Using the optimal values of $\omega$ and $\bb$, compute the solution of equations \eqref{eq:1.1}-\eqref{eq:1.2} for the entire domain.
    %\end{algorithmic}
    \Require Divide the entire domain into non--overlapping rectangular sub--domains and discretize each sub--domain using rectangular grids.
     Use these grid points as inputs. Use previously computed optimal $\omega$ and $\bb$ for the entire domain as the initial weight and bias for each sub--domains network. The boundary conditions for each sub--domains can be obtained from the final trained velocity of the entire domain. For each sub--domain, the following steps are required to generate a sub--domain solution.
         \While {$iter~ <~ epoch$}
         \State  Calculate $[\ub_{\mathcal{N}}^{k}~p_{\mathcal{N}}^{k}]^{T}\approx \z_{K}^{k} ={\omega_{K}^{k}}^{\text{T}}\z_{K-1}^{k}+\bb_{K}^{k}$ using $ \z_{l}^{k}$ where $l=1,\cdots,K-1$ and $k$ denotes the index of each sub--domain.
         \State Reshape the output variables to a matrix form and add the boundary condition of the velocity component.
         \State  Compute derivatives of the differentiable terms involved in the PDEs using the finite difference methods. 
         \State  Calculate the total loss function using \eqref{eq:4} along with the incompressible condition.
         \State Using L--BFGS algorithm, minimize the loss function to obtain the values of $\omega^{k}=\{\omega_{l}^{k} \mid l=1,2,\cdots,K \}$ and $\bb^{k}=\{\bb_{l}^{k} \mid l=1,2,\cdots,K \}$.
         \State $iter=iter+1$
         \EndWhile
         \State Using the optimal values of $\omega^{k}$ and $\bb^{k}$, compute the solution for the sub--domain.
         \end{algorithmic}
\end{algorithm}
\end{minipage}
\end{center}

\section{Results}\label{sec:3}

In this section, we present several numerical results of the Navier--Stokes equations' solutions for the Reynolds numbers Re=400 and Re=1000 in a square lid--driven cavity. The numerical tests were performed utilizing MATLAB R2023a on a system equipped with an 12th Gen Intel(R) Core(TM) i5--12500 3.00 GHz, under the Windows 10 operating system. At first, we consider a square lid--driven cavity $[0,1]\times[0,1]$ and divide it into rectangular grids. In contrast, other types of grids can also be considered, such as staggered grids, but due to complexity, this has not been pursued here. It should be mentioned that the accuracy of the solution is highly dependent on the number of grid points. There is a risk of overfitting when adding more collocation points. However, still no theory is available to help determining the accurate number of collocation points needed to achieve the desired accuracy. Therefore, according to our computational experience and other existing studies \cite{jiang2023}, discretize the entire domain into $100\times100$ grid points. Apart from this, the accuracy of the solution depends on the architecture of the neural networks, especially the number of hidden layers and neurons in each layer. In general, the increase in the number of neurons and layers leads to an improvement in the solution's accuracy, but this process often faces the challenge of overfitting and increases computational complexity when training the weights. Similarly, there is no unified theory on choosing the number of hidden layers and neurons in each layer to optimize solution accuracy. As experienced by several computational trials, we have successfully employed three hidden layers, each with 50 neurons, along with a hyperbolic tangent ($tanh$) function. 
\par The FD--PINNs is implemented by using the L--BFGS algorithm with the setting, MaxIteration: $2\times 10^{5}$, MaxFunctionEvaluations $2\times10^{5}$ and OptimalityTolerance = $10^{-8}$. Optimization is stopped based on which one is achieved first, either MaxIteration or MaxFunctionEvaluations. Apart from this, the optimization can also be stopped when the size of the step is less than the value of the step size tolerance, and constraints are satisfied within the value of optimal tolerance, but this is only happening for few sub--domains cases (approximately 1/3 sub--domains out of total), in this case, one might possibly get the local minima and the remaining sub--domains cases optimization stopped due to the reaching of MaxFunctionEvaluations limit = $2\times 10^{5}$.

\subsection{Reference solution of lid--driven cavity flow for Re=1000}
\begin{figure}[H]
\centering
  \begin{tabular}{@{}ccc@{}}
    \includegraphics[scale=.3]{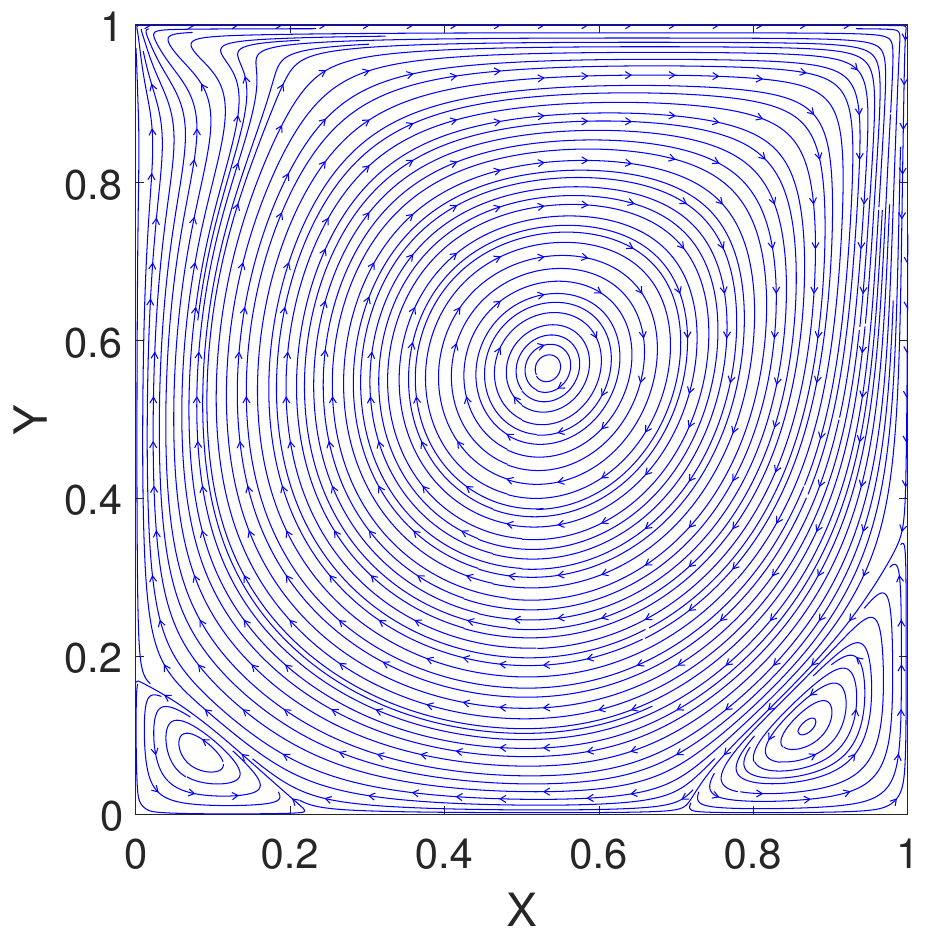} & 
      \includegraphics[scale=.3]{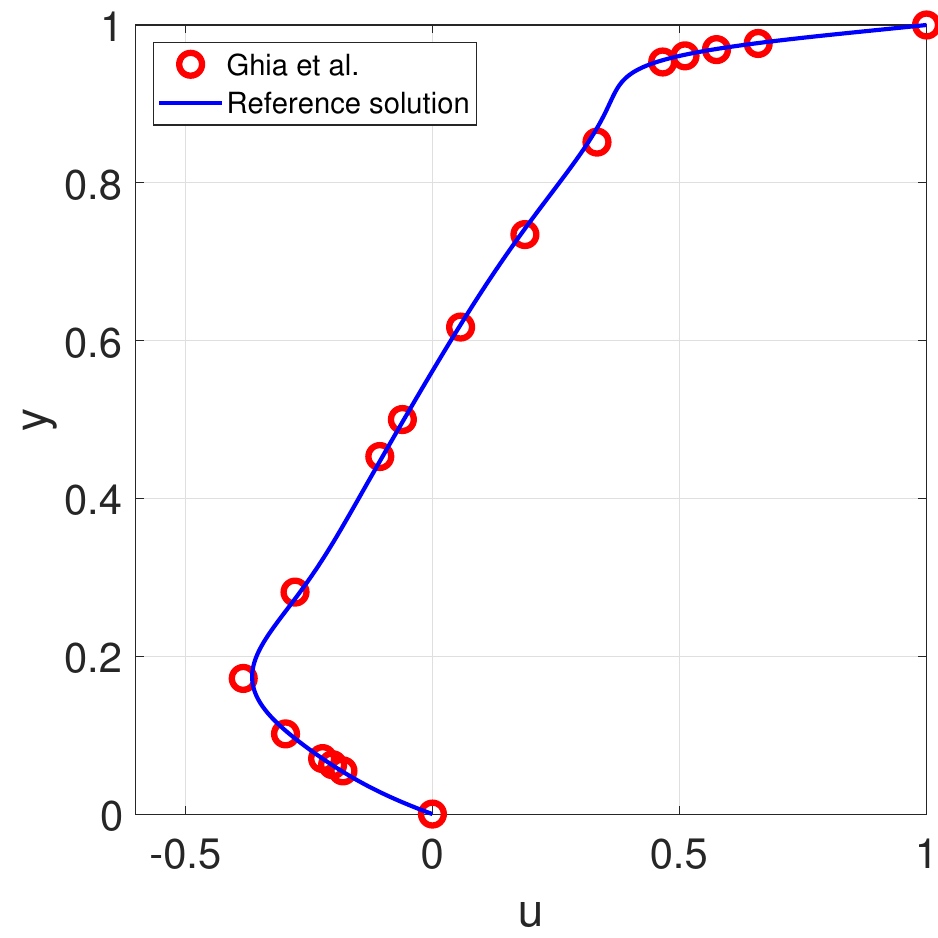} &
      \includegraphics[scale=.3]{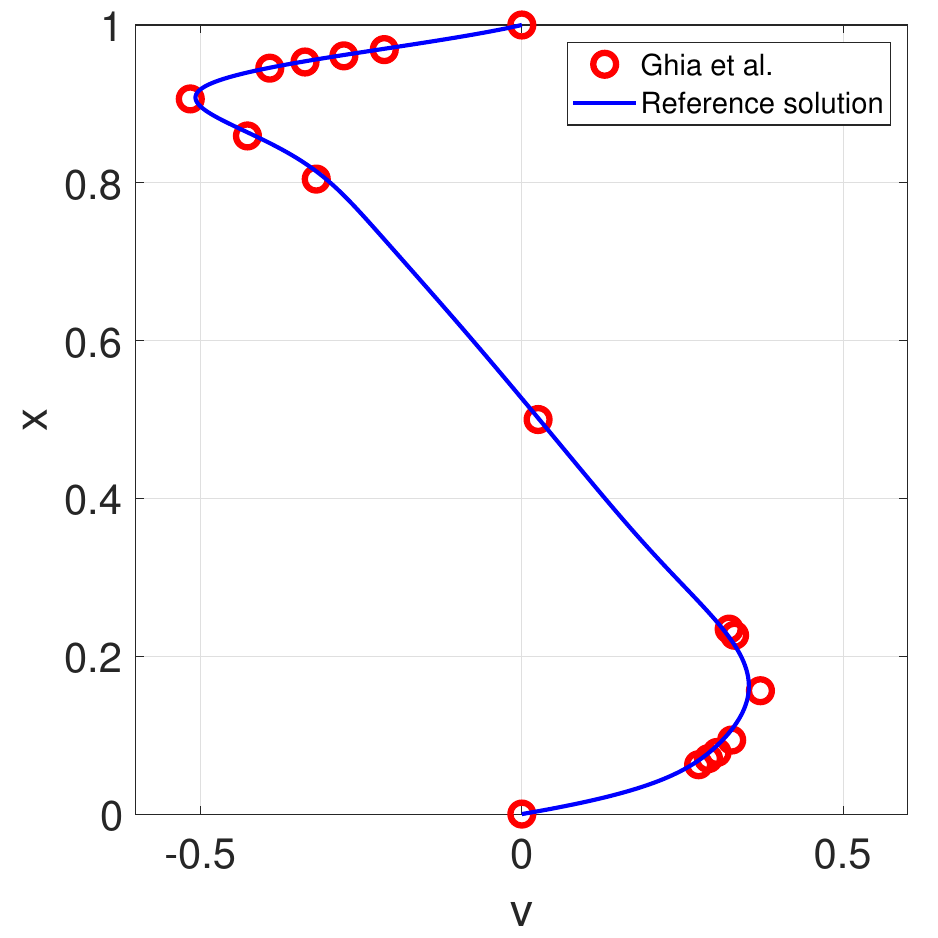} \\
    \textbf{(a)} & \textbf{(b)} & \textbf{(c)} 
  \end{tabular}

   \caption{a.) Fluid streamlines using the reference solution for the Reynolds number Re=1000, b.) Comparison of the horizontal component of the reference velocity along the vertical centerline of the cavity against the Ghia et al. \cite{ghia1982high}, c.) Comparison of the vertical component of the reference velocity along the horizontal centerline of the cavity against the Ghia et al. \cite{ghia1982high}. }
    \label{fig:ref}
\end{figure}

 To check the accuracy of the obtained solution via the present FD--PINNs, we generate the reference solution of the Navier--Stokes equations for Re=1000 in a square lid--driven cavity of size $[0,~1]\times[0,~1]$. We generate the reference solution using the finite difference method described in \cite{griebel1998} with $500\times500$ grid points and successive over-relaxation (SOR) methods applied for pressure Poisson. The reference solution for the Reynolds number Re=1000 is compared with the Ghia et al. \cite{ghia1982high} to check the accuracy of the obtained solution; see Figure \ref{fig:ref}. It can be seen that the horizontal component of the reference velocity along the vertical centerline and the vertical component of the reference velocity along the horizontal centerline match with the solution of Ghia et al. \cite{ghia1982high}. Besides that, the secondary vortices in both corners are exactly similar to those of Ghia et al. \cite{ghia1982high}, and the comparison of the center of both the secondary vortices against the previous literature can be seen in Table \ref{tab:0}.

\subsection{Simulation of lid--driven cavity flow via the standard FD--PINNs for Re=1000}

\begin{figure}[H]
\centering
  \begin{tabular}{@{}cc@{}}
    \includegraphics[scale=.3]{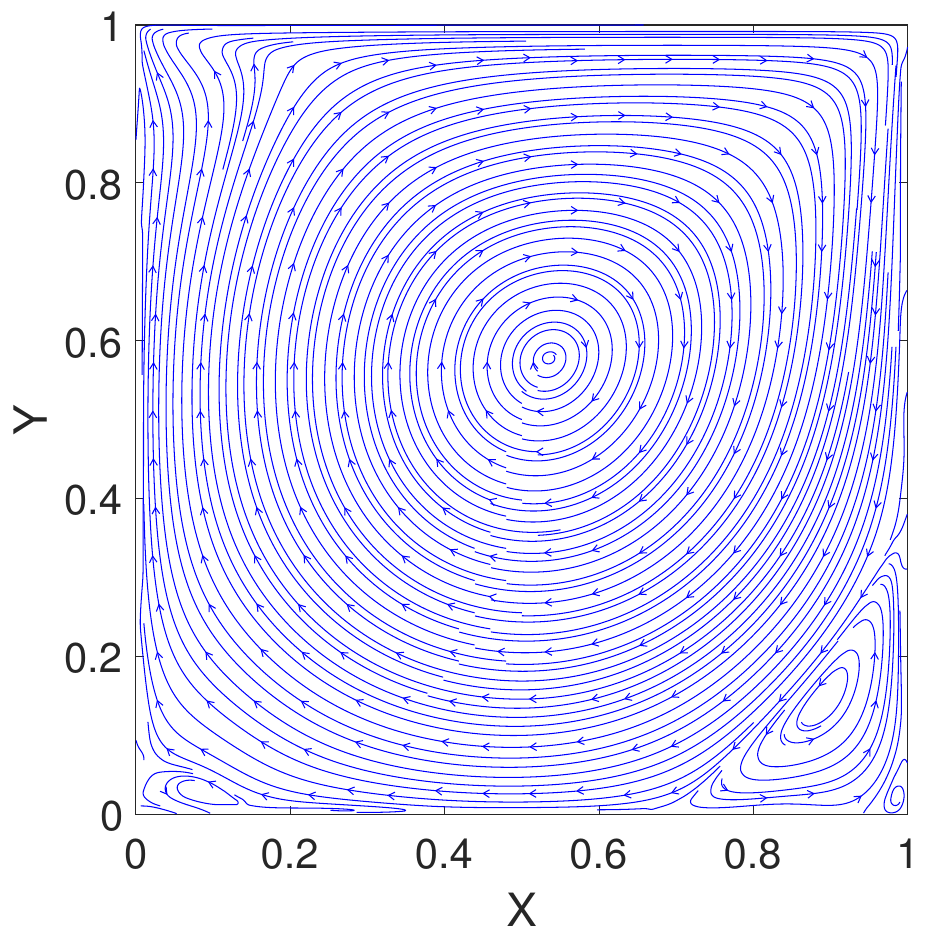} & 
      \includegraphics[scale=.3]{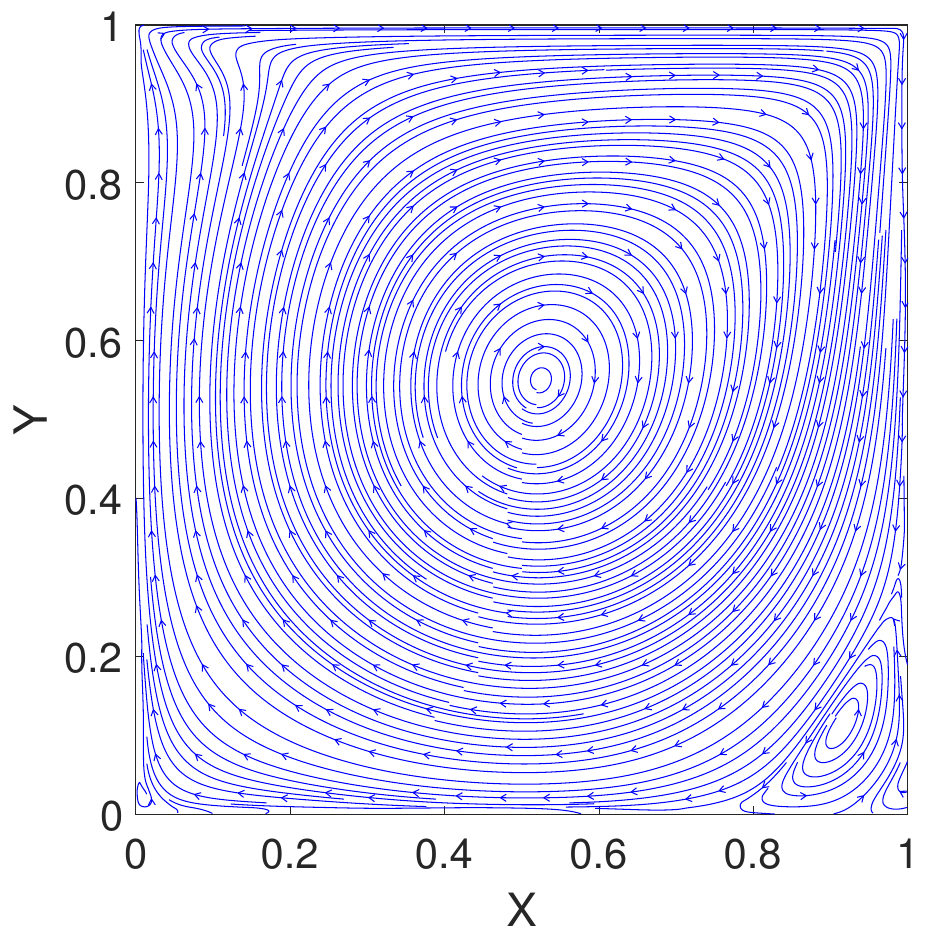} 
\\
    \textbf{(a)} & \textbf{(b)}  \\
   
  \end{tabular}
   \begin{tabular}{@{}cc@{}}
     \includegraphics[scale=.3]{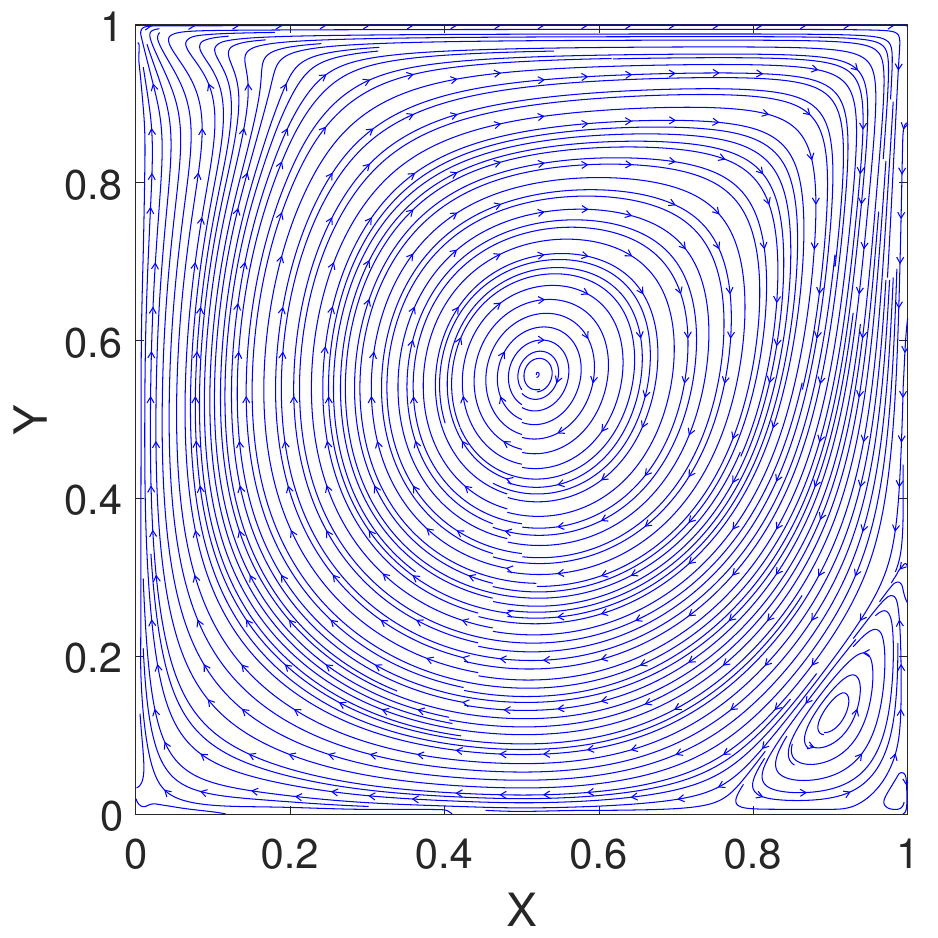} &
   \includegraphics[scale=.3]{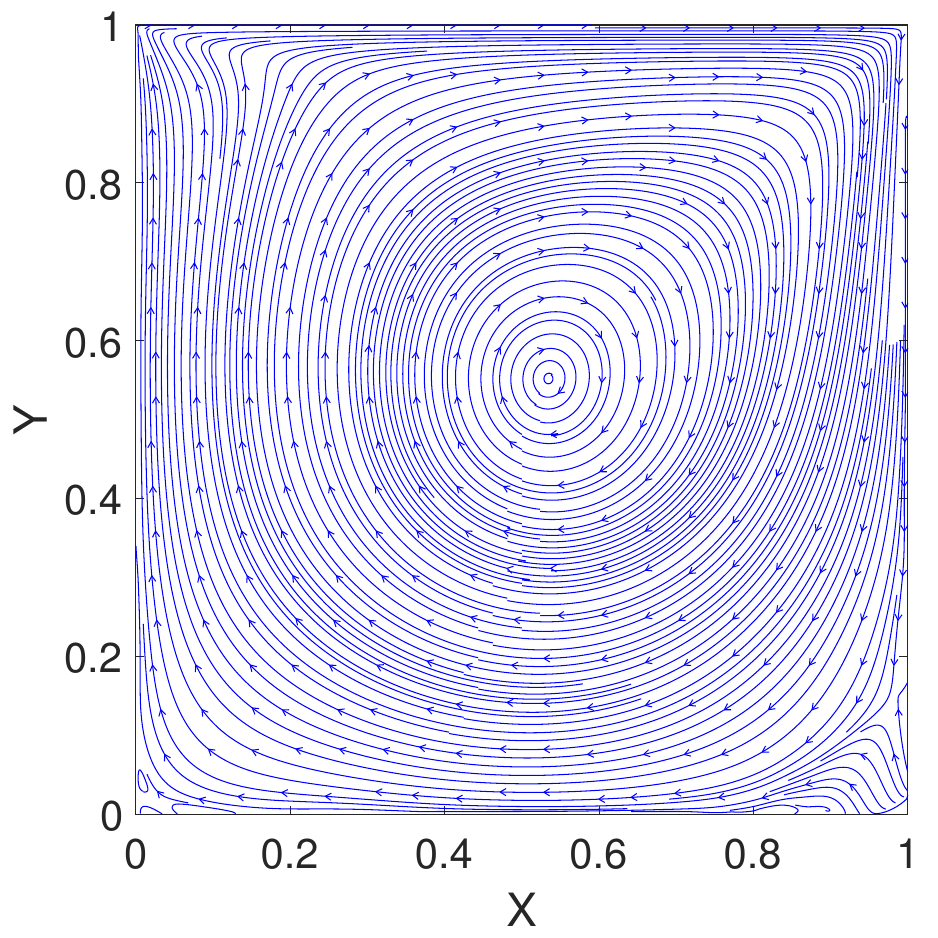} 
 \\
    \textbf{(c)} & \textbf{(d)}  \\

  \end{tabular}

   \caption{Fluid streamlines for the Reynolds number Re=1000 using the standard FD--PINNs with a.) $100 \times 100$ grid points, b.) $120 \times 120$ grid points, c.) $150 \times 150$ grid points, and d.) $300 \times 300$ grid points. Note that the grid points mentioned throughout this section are referred to as the interior grid points of a domain or sub--domain. }
    \label{fig:3}
\end{figure}

The fluid streamlines in the lid--driven cavity for the Reynolds number Re=1000 of the standard FD--PINNs for the $100\times 100$, $120\times 120$, $150\times 150$ and $300\times 300$ grid points are depicted in Figures \ref{fig:3}(a), \ref{fig:3}(b), \ref{fig:3}(c) and \ref{fig:3}(d), respectively.

\begin{figure}[H]
\centering
  \begin{tabular}{@{}c@{}}
    \includegraphics[scale=.5]{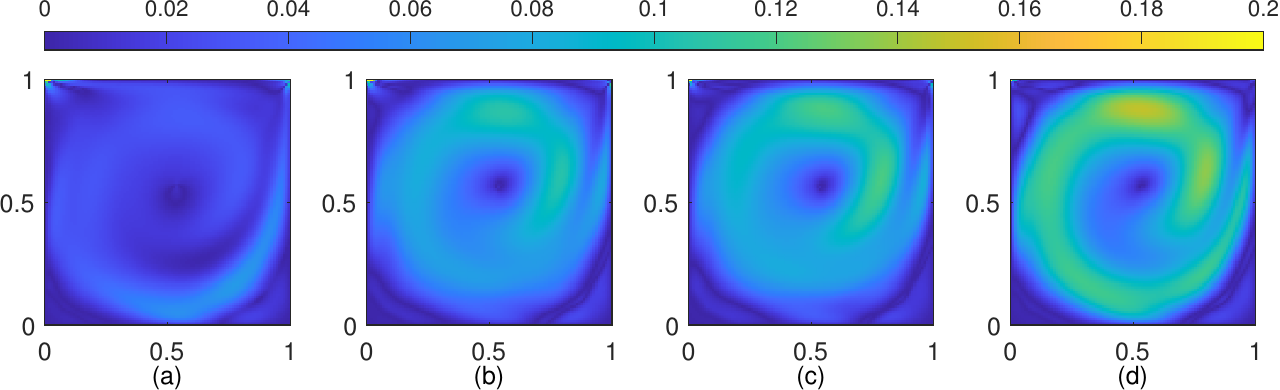} \\
      \includegraphics[scale=.5]{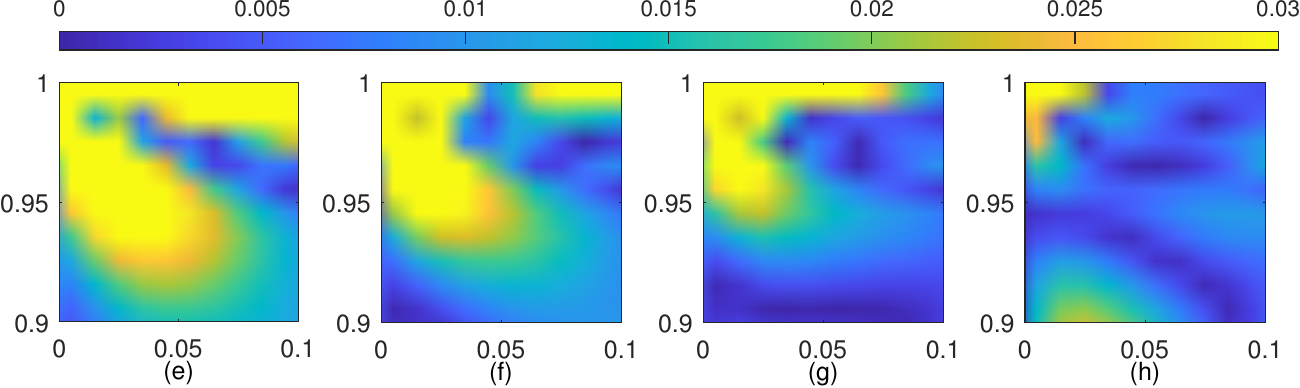} \\
        \includegraphics[scale=.5]{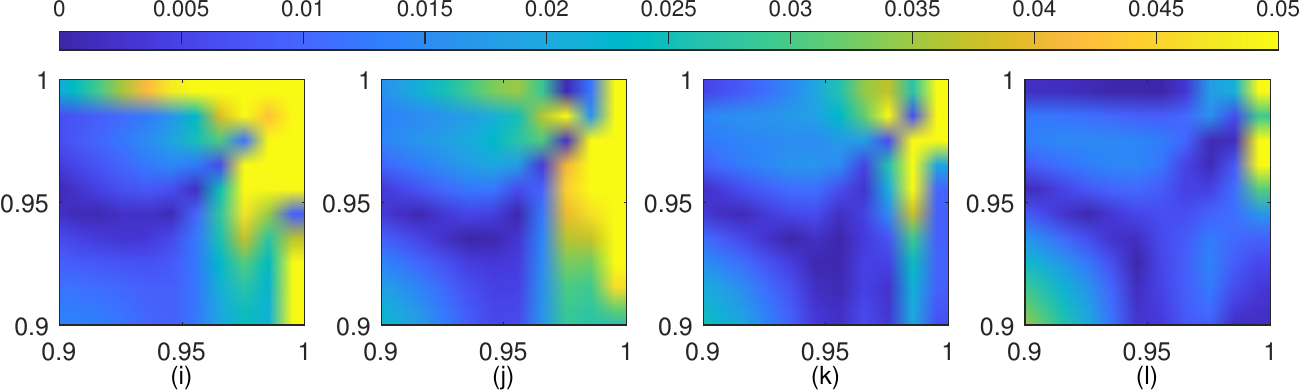} \\
      \includegraphics[scale=.5]{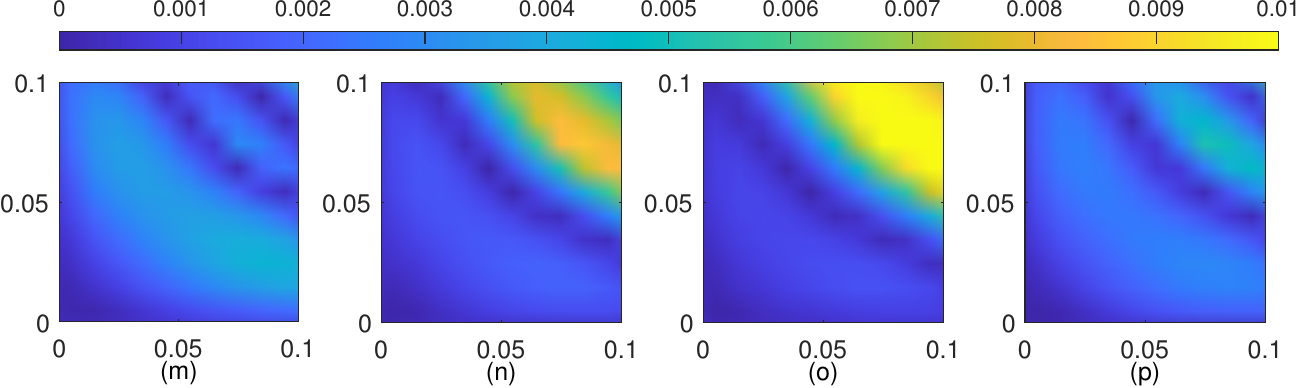} \\
        \includegraphics[scale=.5]{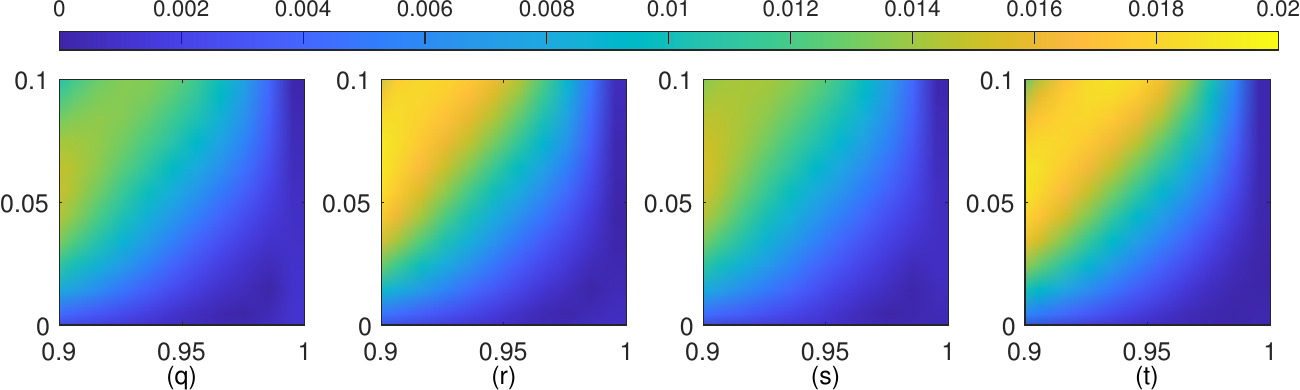} \\ 
  \end{tabular}

   \caption{a--d.) Error in velocity magnitude generated by the standard FD--PINNs for the Reynolds number Re=1000 with $100 \times 100$, $120 \times 120$, $150 \times 150$ and $300 \times 300$ grid points, respectively. e--h.) Error in the lop left corner of the cavity, i--l.) Error in the top right corner, m-p.) Error in the lower left corner, q--t.) Error in the lower right corner.}
    \label{fig:error_1}
\end{figure}

It is seen that a secondary vortex appears in the lower right corner of the lid--driven cavity with low accuracy, while the secondary vortex in the lower left corner almost disappears. Nevertheless, the secondary vortex in the lower left corner of the cavity slightly appears with very low accuracy for the grid points of $100\times 100$. However, the vortex structure is far from the vortex of the reference solution; see Figure \ref{fig:ref}(a), whereas, as the grid points increase, the vortex completely disappears. To understand the effect of the grid points on the solution generated by the standard FD--PINNs, the absolute error of the computed velocity magnitude given by $\left||\ub|-|\ub_{ref}|\right|$ for the four types of grid points mentioned above using the reference solution is shown in Figure \ref{fig:error_1}. It is clearly seen that the error increases for FD--PINNs with very finer grid points; this may be due to the overpredicting behavior of the solution, which arises when neural networks try to capture very small oscillations generated by finite difference discretization for the fine grids. In this case, the neural network model overpredicts instead of forming the true solutions.  On the other hand, an opposite trend is observed for the classical finite difference methods, i.e., finer grid points provide better accuracy as there is no neural network architecture that may try to fix the small oscillation generated by the finite difference discretization and provide overpredicting solutions.

% \begin{figure}[H]
% \centering
%   \begin{tabular}{@{}cc@{}}
%     \includegraphics[width=.48\textwidth]{Streams_500_pinns.pdf} &
%     \includegraphics[width=.48\textwidth]{Streams_500_modi_pinns.pdf} \\
%     \textbf{(a)}  & \textbf{(b)} \\
%   \end{tabular}
%   \caption{a.) Fluid streamlines for the Reynolds number Re=1000 using standard FD--Pinns at the lower right corner of the cavity. b.) Fluid streamlines for the Reynolds number Re=1000 using present technique at the lower right corner of the cavity.}
%   \label{fig:5}
% \end{figure}

\subsection{Simulation of lid--driven cavity flow via the present FD--PINNs for Re=1000}
\begin{figure}[H]
\centering
 \begin{tabular}{@{}cc@{}}
    \includegraphics[scale=.3]{Standard_FDP_100.pdf} & 
      \includegraphics[scale=.3]{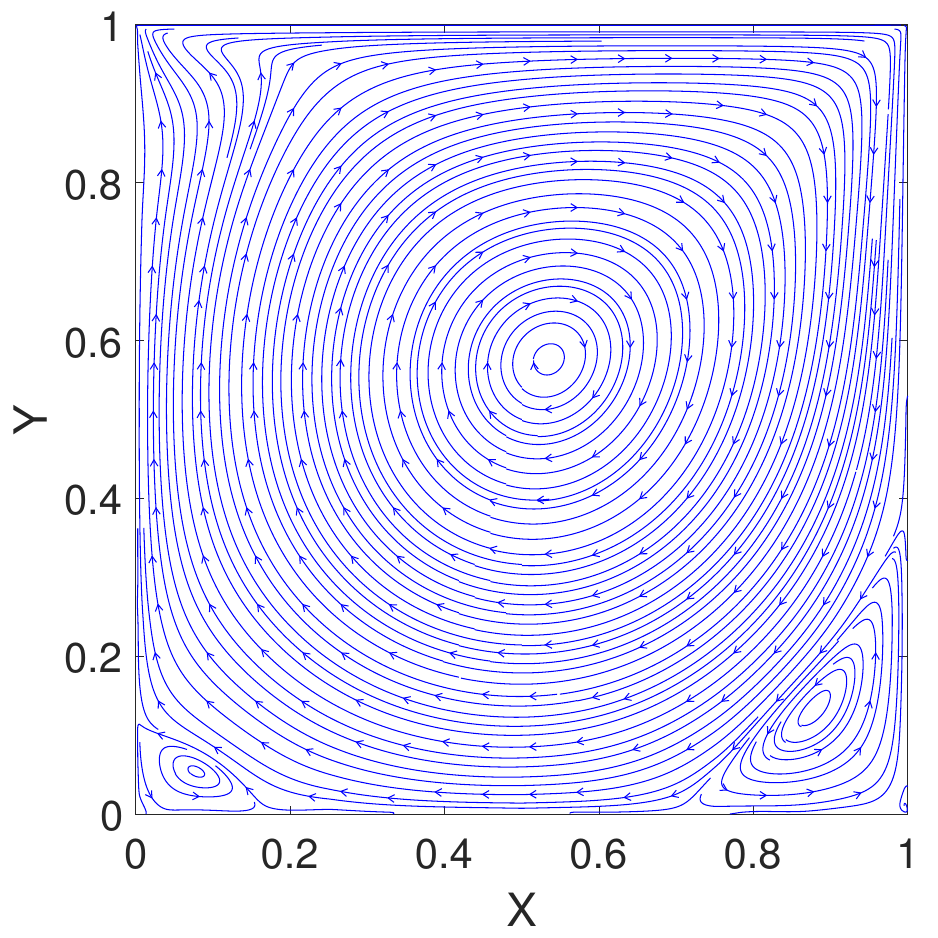} 
\\
    \textbf{(a)} & \textbf{(b)}  \\
   
  \end{tabular}
   \begin{tabular}{@{}cc@{}}
     \includegraphics[scale=.3]{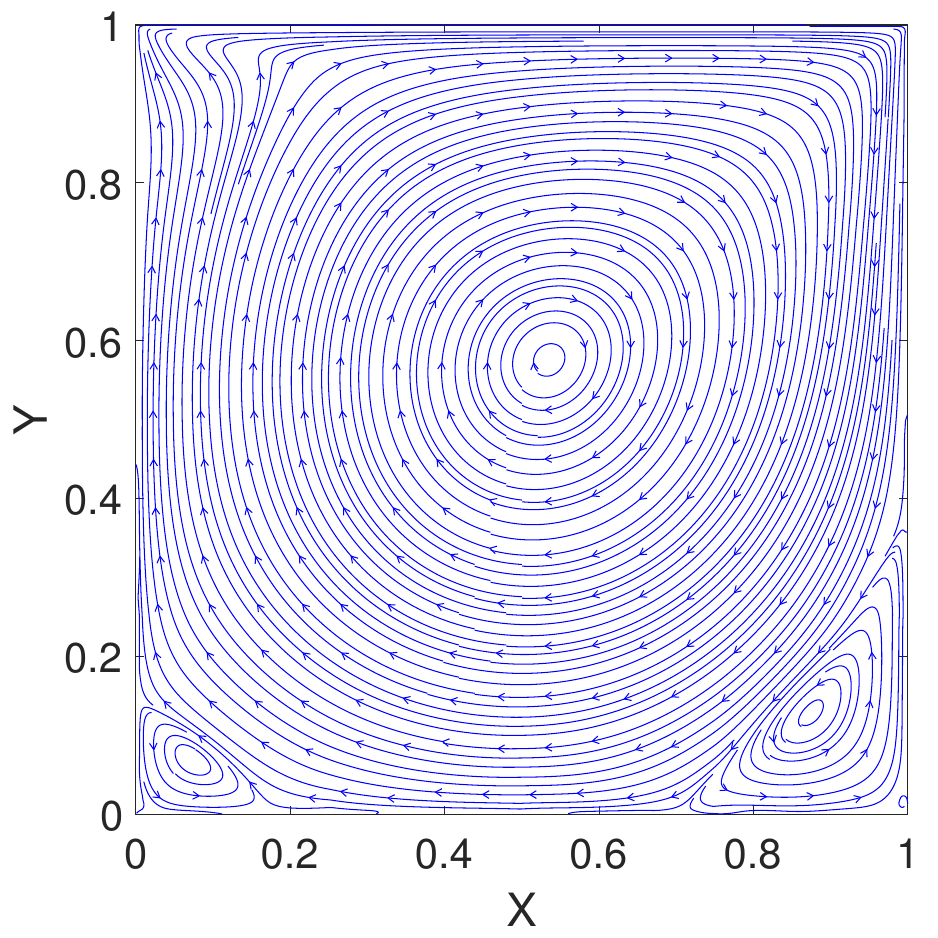} &
   \includegraphics[scale=.3]{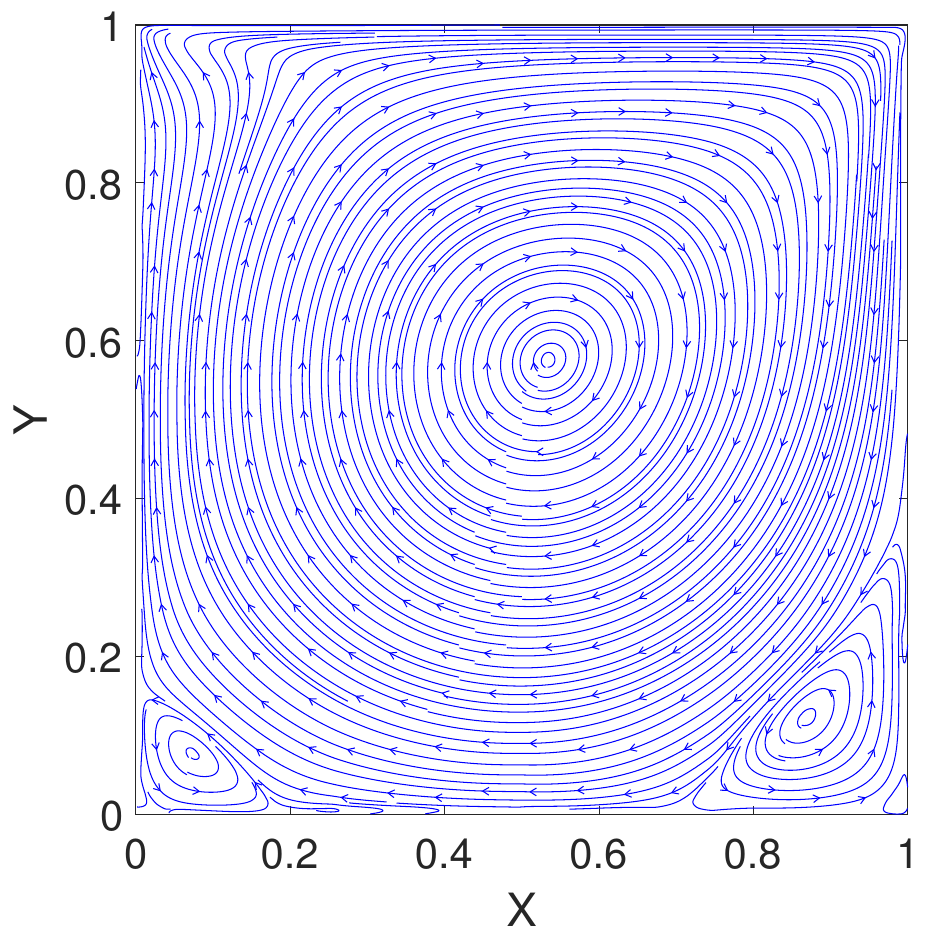} 
 \\
    \textbf{(c)} & \textbf{(d)}  \\

  \end{tabular}
  % \begin{tabular}{@{}cc@{}}
  %   \includegraphics[scale=.4]{Streams_25grid.pdf} &\\
  %  \includegraphics[scale=.35]{Streams_25grid.pdf} \\
   
  %   \textbf{(e)} &   \\
   
  % \end{tabular}
  
   \caption{a.) Fluid streamlines using the standard FD--PINNs with $100\times100$ grid points. Fluid streamlines using the present FD--PINNs with different sub--domain grid points: b.) $100\times100$, c.) $50\times50$, and d.) $25\times25$.}
  \label{fig:4}
\end{figure}

 The technique described in the above section is used to improve the solution accuracy near the walls of the cavity and achieve accurate secondary vortices in the corners. Initially, we have divided the entire domain into 16 sub--domains and discretized each of the sub--domains into three different types of grid points: $100\times 100$, $50\times 50$  and $25\times 25$. The optimization runs with three hidden layers and fifty neurons in each layer, which remain the same for all sub--domains. The streamlines generated by applying the present technique for $100\times 100$ grid points for each sub--domain are shown in Figure \ref{fig:4}(b), and the same for $50\times 50$  and $25\times 25$ grid points are shown in Figures \ref{fig:4}(c) and \ref{fig:4}(d), respectively. It is noticed that the secondary vortex is represented in the lower left corner for all numbers of grid points, which is significantly more accurate than the secondary vortex obtained from the standard FD--PINNs. Moreover, it is match to the classical numerical results \cite{ghia1982high,erturk2005numerical}.

\begin{figure}[H]
\centering
  \begin{tabular}{@{}cccc@{}}
    \includegraphics[scale=.23]{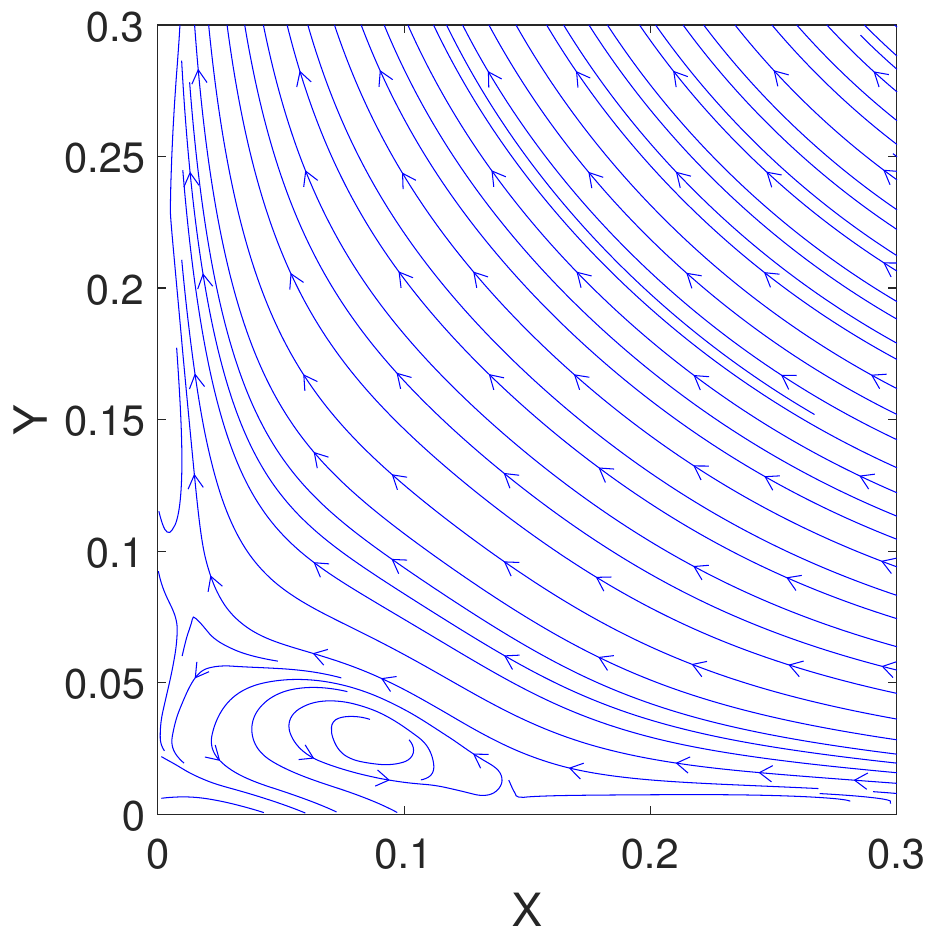} &
    \includegraphics[scale=.23]{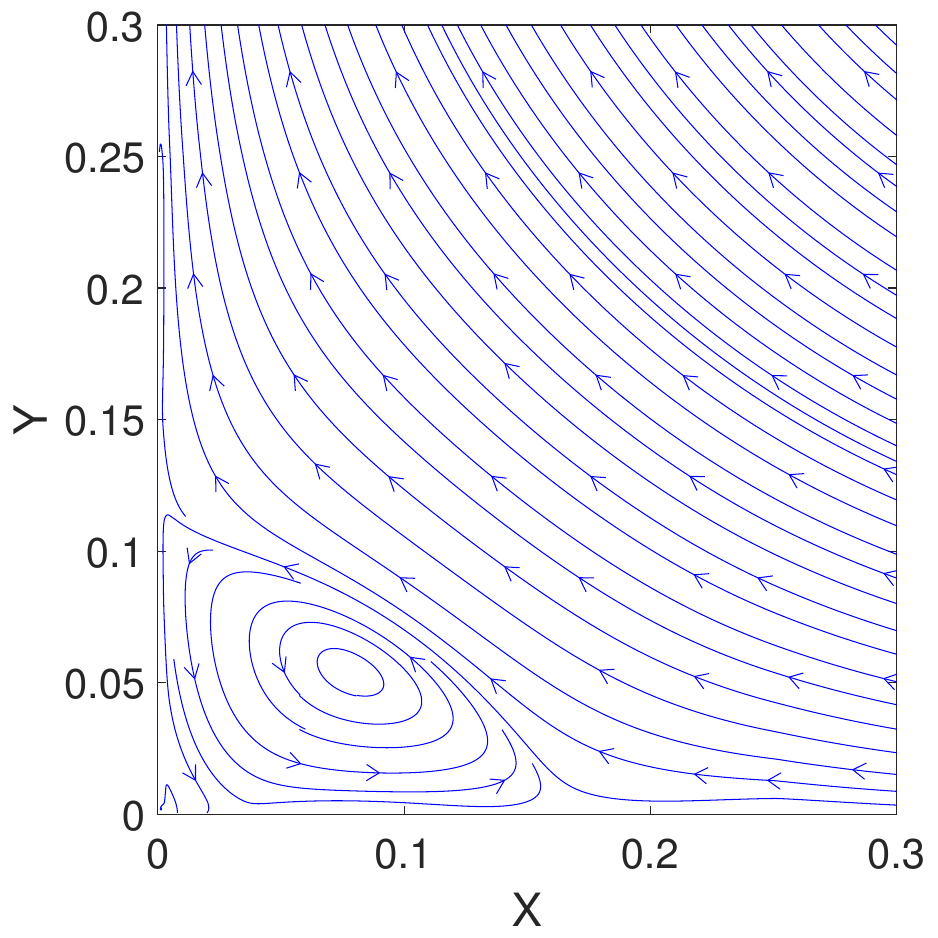} &  \includegraphics[scale=.23]{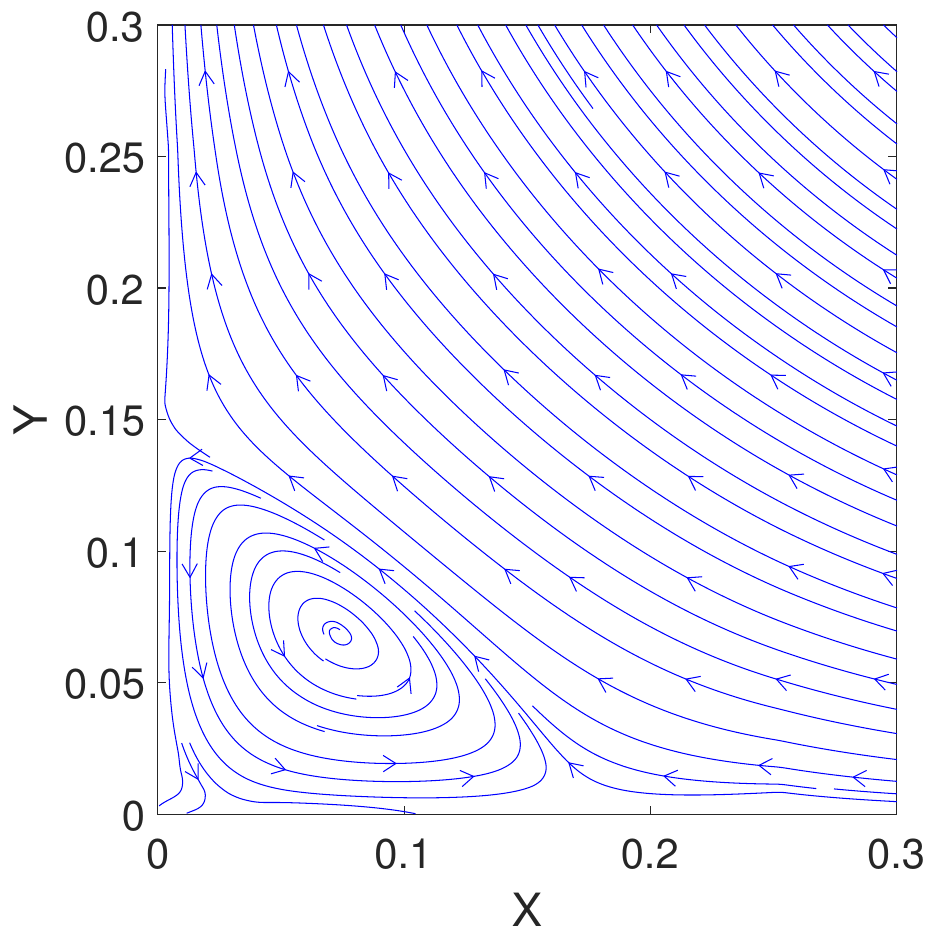} &  \includegraphics[scale=.23]{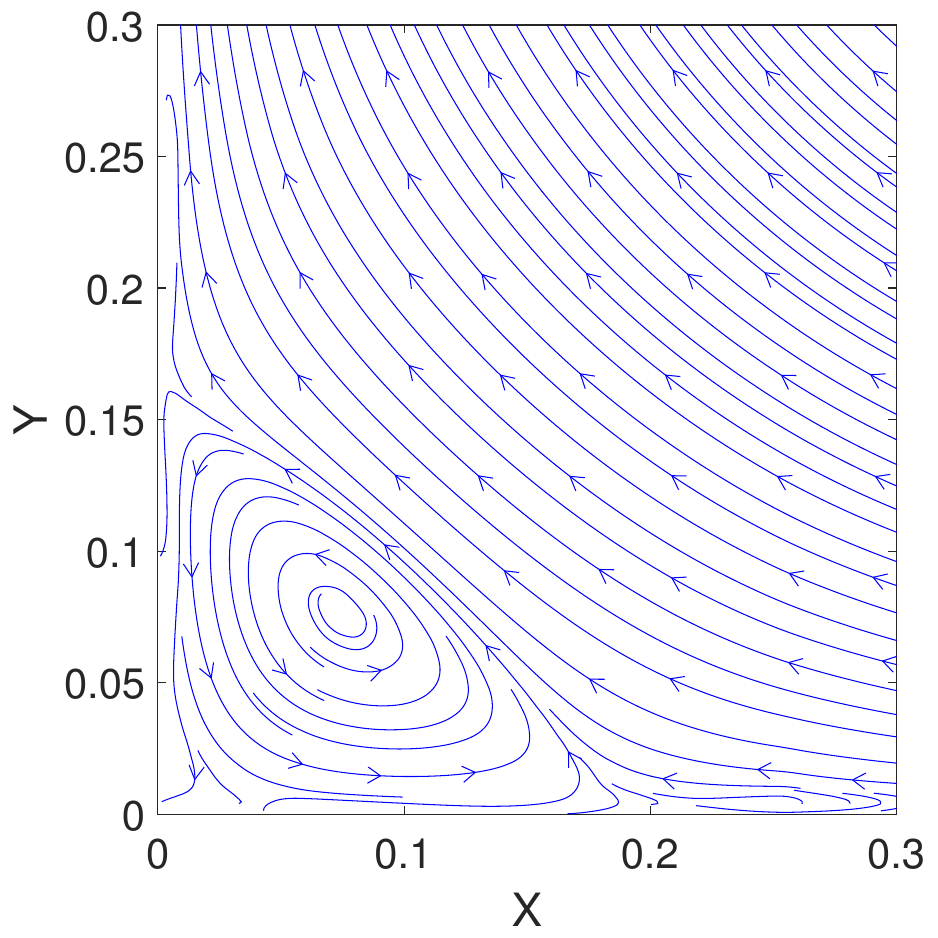} \\
    \textbf{(a)}  & \textbf{(b)} &  \textbf{(c)} & \textbf{(d)} \\
  \end{tabular}
  
   \caption{a.) Fluid streamlines using the standard FD--PINNs with $100\times100$ grid points  in the lower left corner of the cavity. Fluid streamlines using the present FD--PINNs with different sub--domain grid points  in the lower left corner of the cavity: b.) $100\times100$, c.) $50\times50$, and d.) $25\times25$.}
  \label{fig:5}
\end{figure}

\begin{figure}[H]
\centering
  \begin{tabular}{@{}cccc@{}}
    \includegraphics[scale=.23]{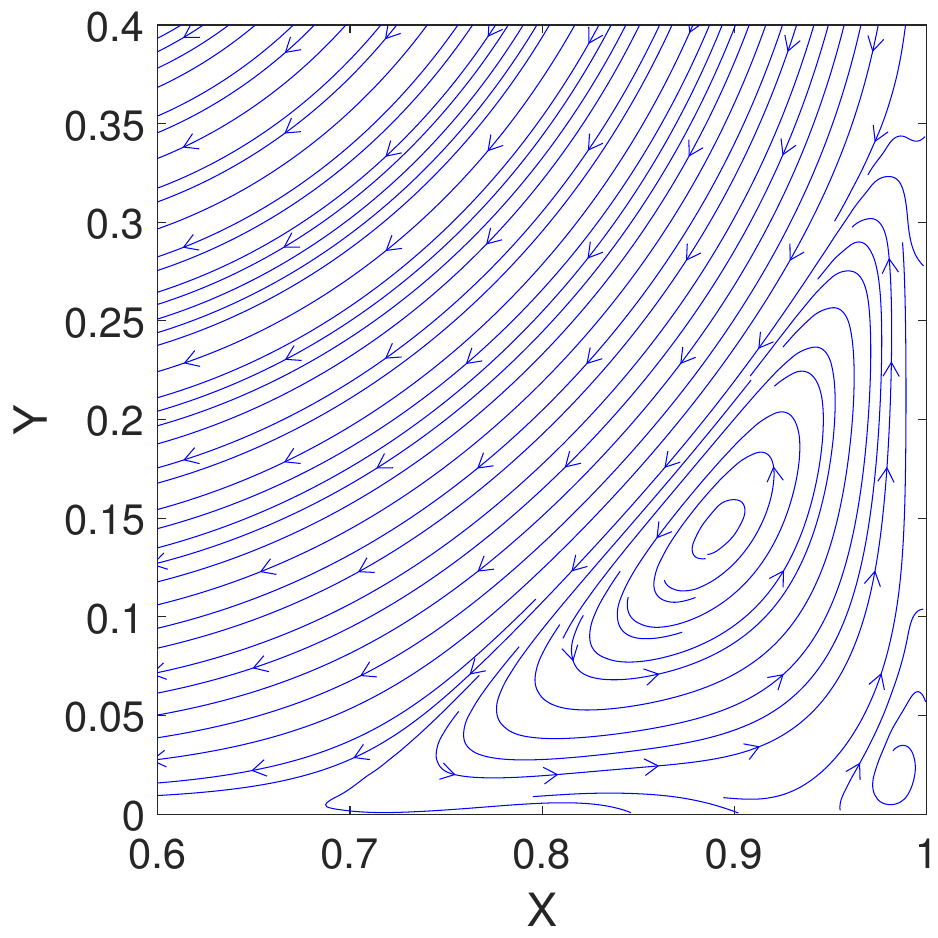} &
    \includegraphics[scale=.23]{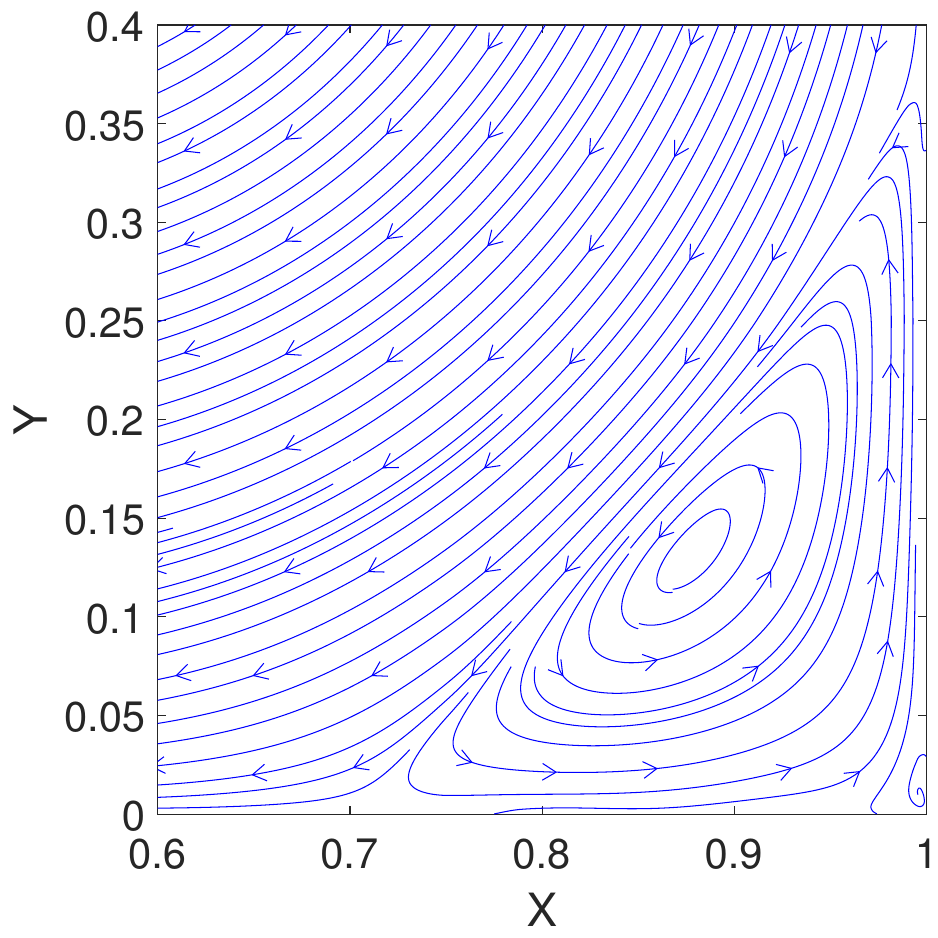} & \includegraphics[scale=.23]{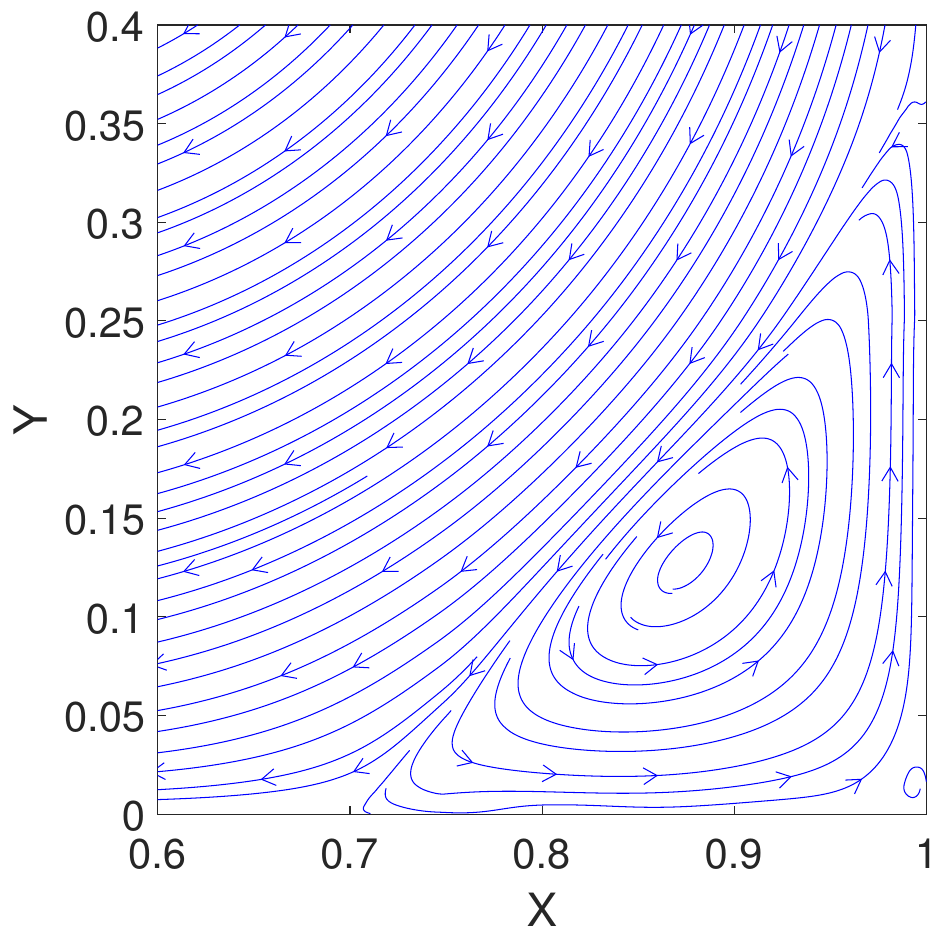} &  \includegraphics[scale=.23]{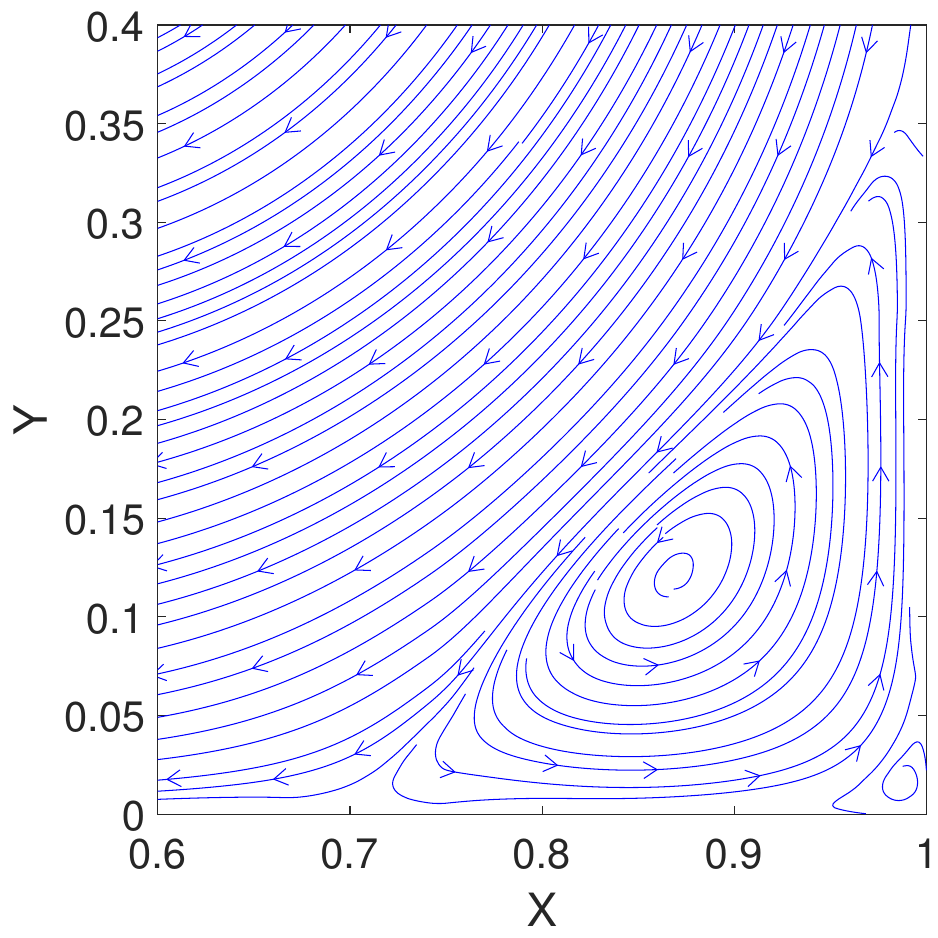}\\
    \textbf{(a)}  & \textbf{(b)} &  \textbf{(c)} & \textbf{(d)} \\
  \end{tabular}
 \caption{a.) Fluid streamlines using the standard FD--PINNs with $100\times100$ grid points  in the lower right corner of the cavity. Fluid streamlines using the present FD--PINNs with different sub--domain grid points  in the lower right corner of the cavity: b.) $100\times100$, c.) $50\times50$, and d.) $25\times25$.}
  \label{fig:6}
\end{figure}

\begin{figure}[H]
\centering
  \begin{tabular}{@{}c@{}}
    \includegraphics[scale=.49]{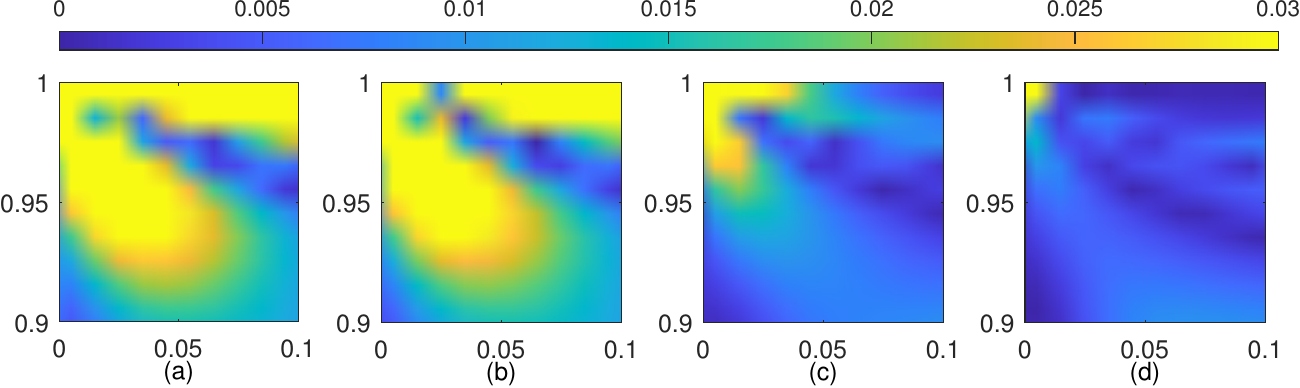} \\
      \includegraphics[scale=.49]{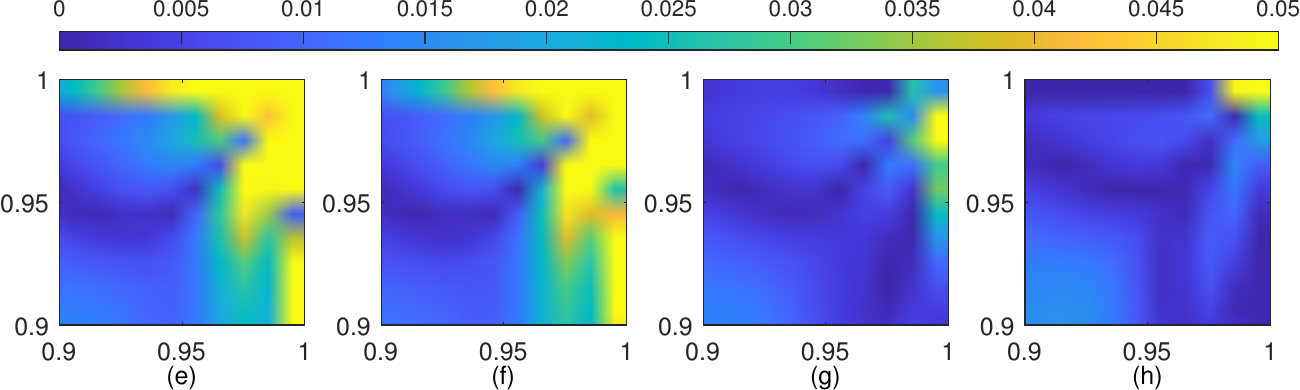} \\
        \includegraphics[scale=.49]{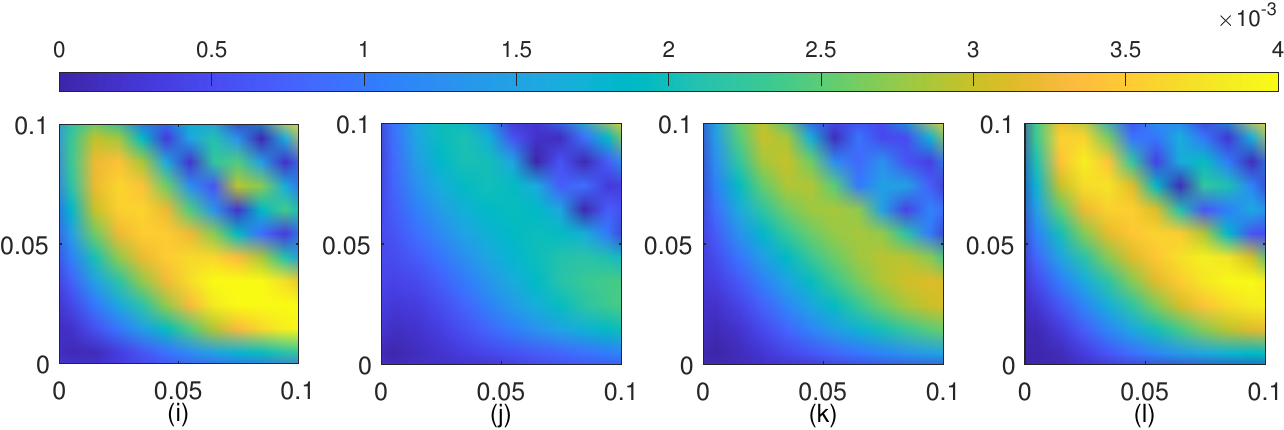} \\
      \includegraphics[scale=.49]{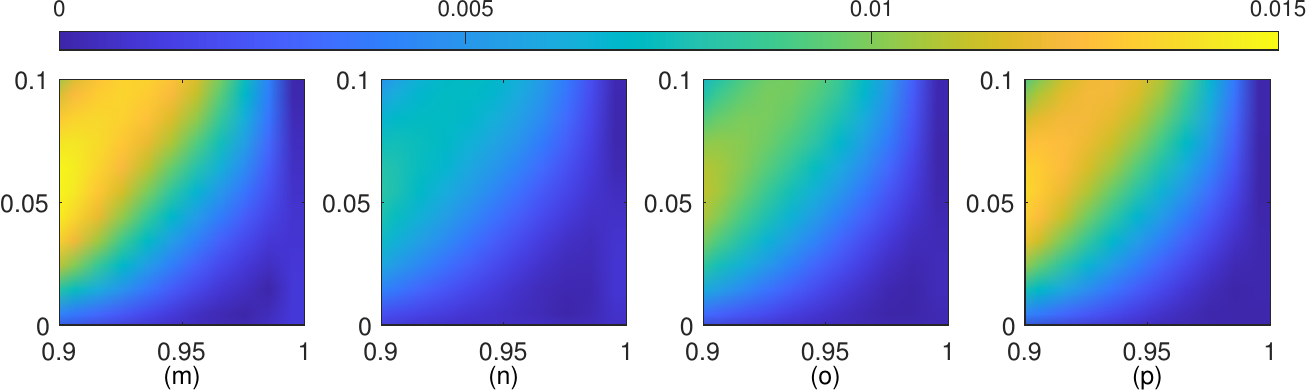} \\
  \end{tabular}

  \caption{a--d.) Error in the velocity magnitude for the Reynolds number Re=1000 generated by the standard FD--PINNs in the top left corner of the cavity with $100 \times 100$ grid points, the present FD--PINNs with $25 \times 25$, $50 \times 50$ and $100 \times 100$ sub--domain grid points, respectively. e--h.) Error in the lop right corner of the cavity, i--l.) Error in the lower left corner, m--p.) Error in the lower right corner.}
    \label{fig:error_11}
\end{figure}

To enhance visual clarity, we have given a pictorial representation of the fluid streamlines for the lower left corner in Figures \ref{fig:5}(a),  \ref{fig:5}(b),  \ref{fig:5}(c) and \ref{fig:5}(d) for the standard FD--PINNs, the present FD--PINNs with $100\times 100$, $50\times 50$ and $25\times 25 $ sub--domain's grid points, respectively. Consistently, the lower right corner of the cavity is shown in Figure \ref{fig:6}. The solution achieved in the lower left corner exceeds that of the standard FD--PINNs in terms of accuracy for all numbers of grid points.  Having a closer look at Figure \ref{fig:6}, clearly understand that although the standard FD--PINNs also produced the secondary vortex in the right corner of the cavity, but it gives less satisfactory results compared to the present FD--PINNs. The absolute error of the velocity magnitude generated by the standard FD--PINNs for $100\times 100$ grid points, the present FD--PINNs for $25 \times 25 $, $50\times 50$,  $100\times 100$ sub--domain grid points using the reference solution is shown in Figure \ref{fig:error_11}. It can be seen that the solution in the lower left and right corners for $25\times 25 $ sub--domain grid points provided less error compared to $50\times 50$ and  $100\times 100$ grid points, whereas a reverse trend is observed for the upper left and right corners. In the region near the lower corners of the cavity, low intensity and highly stable flow is observed compared to the upper corners since it is far from the lid compared to the upper corners. Therefore, a small number of sub--domain grid points is enough to generate an accurate solution near the lower corners of the cavity due to the stable and less non--linear effect of the flow. In contrast, increasing the grid points may lead to overprediction near the lower corners. The location of the centers of the secondary vortices in both lower corners of the cavity for the present FD--PINNs and the previous literature is shown in Table \ref{tab:0}. It is seen that the centers of the secondary vortices generated using the present technique in both the corners of the cavity are in good agreement with the reference values, whereas, for the standard FD--PINNs, the location of the centers is far off from the reference values. Furthermore, the most accurate location of secondary vortices in the lower corners of the cavity are obtained for the case of $25\times25$ sub--domain grid points; the reason behind this phenomenon has already been discussed above.
%\begin{figure}[H]
%\centering
%  \begin{tabular}{@{}cc@{}}
%    \includegraphics[scale=.3]{lower_left_fd_pinns.pdf} &
%    \includegraphics[scale=.3]{lower_left_100.pdf} \\
%    \textbf{(a)}  & \textbf{(b)} \\
%  \end{tabular}
%  \begin{tabular}{@{}ccc@{}}
%     \includegraphics[scale=.3]{lower_left_50.pdf} &
%    \includegraphics[scale=.3]{lower_left_25.pdf} \\
%   
%    \textbf{(c)} & \textbf{(d)}  \\
%   
%  \end{tabular}
%  \caption{a.) Fluid streamlines for the Reynolds number Re=1000 using the standard FD--PINNs with $100\times100$ grid points in the lower left corner of the cavity. b.) Fluid streamlines for the Reynolds number Re=1000 using the present FD--PINNs in the lower left corner of the cavity with $100\times100$ sub--domain grid points. c.) Fluid streamlines for the Reynolds number Re=1000 using the present FD--PINNs in the lower left corner of the cavity with $50\times50$ sub--domain grid points. d.) Fluid streamlines for the Reynolds number Re=1000 using the present FD--PINNs in the lower left corner of the cavity with $25\times25$ sub--domain grid points. }
%  \label{fig:5}
%\end{figure}

\begin{table}[H]
 \centering
 
 \caption{\label{tab:0} Comparison of the center of secondary vortices.}
\vspace*{3mm}
 \renewcommand\arraystretch{1.1}
 \resizebox{\textwidth}{!}{
 
 \begin{tabular}{ccccc}
 
  \hline
& \multicolumn{2}{c}{\hspace{1cm} Lower left corner}{\hspace{1cm}} & \multicolumn{2}{c}{Lower right corner}\\
\cline{2-5}
  Source     &     x   &  y &  x & y \\  
                                                                        
     \hline 
   The present FD--PINNs (25 $\times$ 25 sub--domain grid points)&   0.0735   & 0.0753  &    0.8695   &  0.1210  \\
      
    The present  FD--PINNs (50 $\times$ 50 sub--domain grid points)&   0.0731  & 0.0673   &   0.8750 &  0.1270    \\
     
     The present    FD--PINNs (100 $\times$ 100 sub--domain grid points)&   0.0782  &  0.0535 &    0.8795   &  0.1330     \\
     
     The standard FD--PINNs (100 $\times$ 100 grid points) &   0.0845  & 0.0249   &  0.8940   &  0.1470  \\
     
       Reference solution   &    0.0830   &  0.0780 &   0.8690 & 0.1110   \\
         Botella and Peyret    \cite{botella1998}&  0.0833  &  0.0781 &   0.8640 & 0.1118   \\
        Ghia et al. \cite{ghia1982high}  &    0.0859   &  0.0781 &   0.8594 & 0.1094   \\
      
        Gupta and Kalita \cite{gupta2005new}     & 0.0875   &  0.0750  &   0.8625 & 0.1125  \\

            Bruneau and Jouron \cite{bruneau1990} &   0.0859  & 0.0820  &  0.8711 & 0.1094 \\
         
            Tian and Yu \cite{tian2011efficient}&   0.0859    &    0.0781 &  0.8594 & 0.1094   \\
 Li et al. \cite{li2023non} &   0.0855  & 0.0797 &  0.8619 & 0.1125   \\
\hline
\end{tabular}}

\end{table}

For the details of the solution accuracy, the means square error (MSE) of the horizontal velocity $u$ and vertical velocity $v$, which are defined by $\text{MSE}_{u}=\sum_{i=1}^{N}\frac{1}{N}(u-u_\text{ref})^2$ and $\text{MSE}_{v}=\sum_{i=1}^{N}\frac{1}{N}(v-v_\text{ref})^2$, respectively, is discussed using the reference solution. In Table \ref{tab:1}, we have shown the $\text{MSE}_u$ at nine equidistant vertical lines in the domain $[0.1,0.9]\times[0,1]$ of the cavity. It can be seen that for half of the vertical lines, the present FD--PINNs provides better accuracy than the standard FD--PINNs, whereas a reverse trend is observed for the remaining vertical lines. The present FD--PINNs yield more accurate results for the solutions along the vertical lines in the domains $[0.01,0.09]\times[0,1]$ and $[0.91,.99]\times[0,1]$ of the cavity compared to the standard FD--PINNs, which are reflected in Table \ref{tab:2} and Table \ref{tab:3}, respectively. From Table \ref{tab:2}, it can be observed that the $100\times100$ grid points solutions are slightly better than the $50\times 50$ grid points solutions, while, from Table \ref{tab:3}, mixed results can be seen, i.e., $50\times50$ sub--domains grid points provide better solutions than $100\times100$ grid points solutions along some vertical lines and the remaining vertical lines show the opposite trend. In Table \ref{tab:4}, the $\text{MSE}_v$ in nine equidistant horizontal lines in the domain $[0,1]\times[0.1,0.9]$ of the cavity have been shown. It is worth emphasizing that for most of the horizontal lines the present FD--PINNs is better than the standard FD--PINNs in terms of accuracy. The solutions' accuracy is improved significantly for the horizontal lines in the domains $[0,1]\times[0.01,0.09]$ and $[0,1]\times[0.91,.99]$ of the cavity, which are seen in Table \ref{tab:5} and Table \ref{tab:6}, respectively. It is noticed that in the region near the bottom wall, the $25\times 25$ grids provided a better solution. As increasing the distance of the horizontal line from the bottom wall, the solution accuracy decreases for the grid points  $25\times 25$; see Table \ref{tab:5}. The reverse trend is observed for the horizontal line in the region near the upper walls, i.e., comparably higher grid points provided more accuracy compared to the lower grid points; see Table \ref{tab:6}. The intensity of the velocity along the horizontal line near the bottom wall is very low and highly stable. Therefore, a small number of grid points is enough to generate accurate results in the region near the bottom wall, whereas higher grid points provide the overpredicting behavior of the solution. In other case, to generate an accurate solution in the region near the upper wall where the high intensity and nonlinear flow is present, a slightly higher grid point is required, and in this case, low grid points failed to generate accurate solutions.

\begin{table}[H]
 \centering

 \caption{\label{tab:1} The MSE$_u$  comparison between the present and the standard FD--PINNs for the velocity $u$ at several vertical lines in the domain $[0.1,0.9]\times[0,1]$ of the cavity for the Reynolds number $Re=1000$.}
 \vspace*{3mm}
\renewcommand\arraystretch{1.1}
\resizebox{\textwidth}{!}{
\begin{tabular}{ccccc}
\hline
$x$       &  The present FD--PINNs       &  The present FD--PINNs & The present FD--PINNs                        &  The standard FD--PINNs     \\                                                                           
&($25\times25$ sub--domain grid points )  & ($50\times50$ sub--domain grid points ) & ($100\times100$ sub--domain grid points ) & ($100\times100$ grid points)

       \\\hline
       0.10      &   1.2580e-04   &  4.5056e-05        &   4.2668e-05 &     1.1368e-04    \\
      
        0.20    &    2.8696e-04   & 2.1169e-04        & 1.7691e-04 &      1.7548e-04
    \\
     
       0.30    &     3.5200e-04    & 3.3698e-04        &     3.2870e-04 &      3.8018e-04     \\
     
       0.40     &   6.0982e-04  & 5.6828e-04         &      5.4786e-04 &        5.8704e-04    \\
       
       0.50  &     7.3985e-04    &   7.2464e-04     &   7.1864e-04 &      7.2638e-04   \\
      
       0.60    &    6.9562e-04    & 7.1852e-04       &  7.2616e-04 &      6.7338e-04    \\
        
          0.70  &      4.2846e-04    & 5.0746e-04       &    5.4958e-04 &      4.4032e-04  \\
       
          0.80    &    3.4818e-04    &  3.9585e-04       &  4.2006e-04 &      2.2768e-04   \\
         
          0.90  &        6.2537e-05    &  1.1024e-04       &       1.4756e-04 &         6.1209e-05   \\
     
\hline
\end{tabular}}

\end{table}

\begin{table}[H]
\centering

\caption{\label{tab:2} The MSE$_u$ comparison between the present and the standard FD--PINNs for the velocity $u$ at several vertical lines in the domain $[0.01,0.09]\times[0,1]$ of the cavity for the Reynolds number $Re=1000$.}
\vspace*{3mm}
\renewcommand\arraystretch{1.1}
\resizebox{\textwidth}{!}{
\begin{tabular}{ccccc}
\hline
$x$       &  The present FD--PINNs       &  The present FD--PINNs & The present FD--PINNs                        &  The standard FD--PINNs     \\                                                                           
&($25\times25$ sub--domain grid points )  & ($50\times50$ sub--domain grid points ) & ($100\times100$ sub--domain grid points ) & ($100\times100$ grid points)
 \\ \hline
       
       0.01     & 1.9000e-03   &  1.6973e-04      &    5.3916e-05 &      2.5000e-03    \\
      
        0.02    &    3.6008e-04  &  2.4029e-05      &    2.3642e-06 &     6.9050e-04    \\
     
       0.03    &   6.6109e-05  &   2.0272e-05    &     1.7614e-06 &     1.1031e-04    \\
     
       0.04     & 1.1037e-04  &   1.4707e-05     &        3.7727e-06 &       6.4359e-05    \\
       
       0.05  &      1.6296e-04   &   1.4325e-05     &      7.4290e-06 &        1.2288e-04
   \\
      
       0.06    &  1.7529e-04 &    1.7112e-05   &   1.2526e-05 &        1.5543e-04
   \\
        
          0.07  &      1.6507e-04 &     2.1920e-05     &       1.8831e-05              &      1.5596e-04   \\
       
          0.08    &   1.4893e-04   &    2.8301e-05    &            2.6081e-05 &      1.4224e-04   \\
         
          0.09  &      1.3486e-04   &    3.6040e-05      &          3.4062e-05 &      1.2645e-04   \\
     
\hline
\end{tabular}}

\end{table}

   \begin{table}[H]
   \centering
   
   \caption{\label{tab:3} The MSE$_u$  comparison between the present and the standard FD--PINNs for the velocity $u$ at several vertical lines in the domain $[0.91,0.99]\times[0,1]$ of the cavity for the Reynolds number $Re=1000$.}
   \vspace*{3mm}
\renewcommand\arraystretch{1.1}
\resizebox{\textwidth}{!}{
\begin{tabular}{ccccc}

\hline
$x$       &  The present FD--PINNs       &  The present FD--PINNs & The present FD--PINNs                        &  The standard FD--PINNs     \\                                                                           
&($25\times25$ sub--domain grid points )  & ($50\times50$ sub--domain grid points ) & ($100\times100$ sub--domain grid points ) & ($100\times100$ grid points)
                         
    \\ \hline
    0.91      &   5.2390e-05   & 8.5608e-05     &  1.1784e-04 &    6.2531e-05   \\
      
        0.92    &    5.1543e-05   &  6.4429e-05   &   9.0233e-05 &       7.4399e-05    \\
     
       0.93    &      6.3472e-05    &  4.7954e-05     &  6.5841e-05 &    1.0014e-04     \\
     
       0.94     &      9.2744e-05    &  3.7542e-05     &      4.5487e-05 &     1.4368e-04  \\
       
       0.95  &        1.4582e-04    &  3.4622e-05    &     2.9523e-05 &      2.1055e-04  \\
      
       0.96    &   2.3412e-04    & 4.0939e-05      &   1.7890e-05              &        3.1160e-04   \\
        
          0.97  &     3.8481e-04   &  6.0207e-05     &      1.0457e-05         &      4.7553e-04
    \\
       
          0.98    &     6.7765e-04   &   1.0414e-04       &        7.8566e-06 &      7.8704e-04
   \\
         
          0.99  &      1.3000e-03      &  2.1098e-04      &        1.7036e-05
 &       1.4000e-03   \\
\hline
\end{tabular}}

\end{table}

 \begin{table}[H]
 \centering
 
 \caption{\label{tab:4}The MSE$_v$ comparison between the present and the standard FD--PINNs for the velocity $v$ at several horizontal lines in the domain $[0,1]\times[0.1,0.9]$ of the cavity for the Reynolds number $Re=1000$.}
 \vspace*{3mm}
\renewcommand\arraystretch{1.1}
\resizebox{\textwidth}{!}{
\begin{tabular}{ccccc}
\hline
$y$       &   The present FD--PINNs       &  The present FD--PINNs & The present FD--PINNs                        &  The standard FD--PINNs     \\                                                                           
&($25\times25$ sub--domain grid points )  & ($50\times50$ sub--domain grid points ) & ($100\times100$ sub--domain grid points ) & ($100\times100$ grid points)

                                                                                \\ \hline
       
       0.10      &    2.2068e-04 & 2.0980e-04       & 2.1148e-04 &    2.2793e-04  \\
      
        0.20    &    7.1743e-04  & 5.6499e-04        &   4.9633e-04 &     5.0475e-04    \\
     
       0.30    &   5.4955e-04   & 5.1492e-04       &    4.9534e-04 &    5.8147e-04    \\
     
       0.40     &    7.1272e-04   &  6.3413e-04        &   5.9404e-04
 &       6.5140e-04    \\
       
       0.50  &    7.6007e-04     &   6.8193e-04     &   6.4887e-04
               &        6.8080e-04   \\
      
       0.60    &      5.7655e-04  & 5.6974e-04       &     5.6812e-04 &       5.7609e-04    \\
        
          0.70  &    3.7940e-04   &   4.0006e-04
       &  4.1461e-04 &      3.7729e-04    \\
       
          0.80    &      2.1730e-04   & 2.1654e-04     &         2.1934e-04 &      1.5156e-04
   \\
         
          0.90  &        5.4900e-05   &  3.7537e-05       &        5.0653e-05 &         6.0027e-05
   \\
     
\hline
\end{tabular}}

\end{table}

  \begin{table}[H]\label{tab:5}
  \centering
  
  \caption{\label{tab:5}The MSE$_v$ comparison between the present and the standard FD--PINNs for the velocity $v$ at several horizontal lines in the domain $[0,1]\times[0.01,0.09]$ of the cavity for the Reynolds number $Re=1000$.}
  \vspace*{3mm}
\renewcommand\arraystretch{1.1}
\resizebox{\textwidth}{!}{
\begin{tabular}{ccccc}
\hline
$y$       &   The present FD--PINNs       &  The present FD--PINNs & The present FD--PINNs                        &  The standard FD--PINNs     \\                                                                           
&($25\times25$ sub--domain grid points )  & ($50\times50$ sub--domain grid points ) & ($100\times100$ sub--domain grid points ) & ($100\times100$ grid points)

                                                                                \\ \hline
       
       0.01      &  5.3276e-08  & 5.3765e-08      &    4.4931e-08 &      6.9460e-08   \\
      
        0.02    &   1.0930e-06  &  1.1612e-06     &    1.3284e-06 &    1.2693e-06    \\
     
       0.03    &   5.9607e-06   &   6.1540e-06     &     6.9235e-06 &      7.0389e-06   \\
     
       0.04     &    1.7196e-05  &   1.7545e-05      &      1.9367e-05 &        2.0372e-05
    \\
       
       0.05  &      3.5943e-05   &  3.6378e-05 &    3.9491e-05              &        4.2277e-05
   \\
      
       0.06    &   6.2047e-05   &  6.2281e-05         & 6.6581e-05 &       7.1912e-05
  \\
        
          0.07  & 9.4607e-05    &   9.4023e-05         &   9.9039e-05
 &       1.0737e-04
   \\
       
          0.08    & 1.3252e-04    &   1.3009e-04          &    1.3505e-04             &       1.4645e-04   \\
         
          0.09  &  1.7481e-04    &   1.6908e-04        &     1.7297e-04 &         1.8717e-04   \\
     
\hline
\end{tabular}}

\end{table}

\begin{figure}[H]
\centering
   \begin{tabular}{@{}cccc@{}}
    \includegraphics[scale=.23]{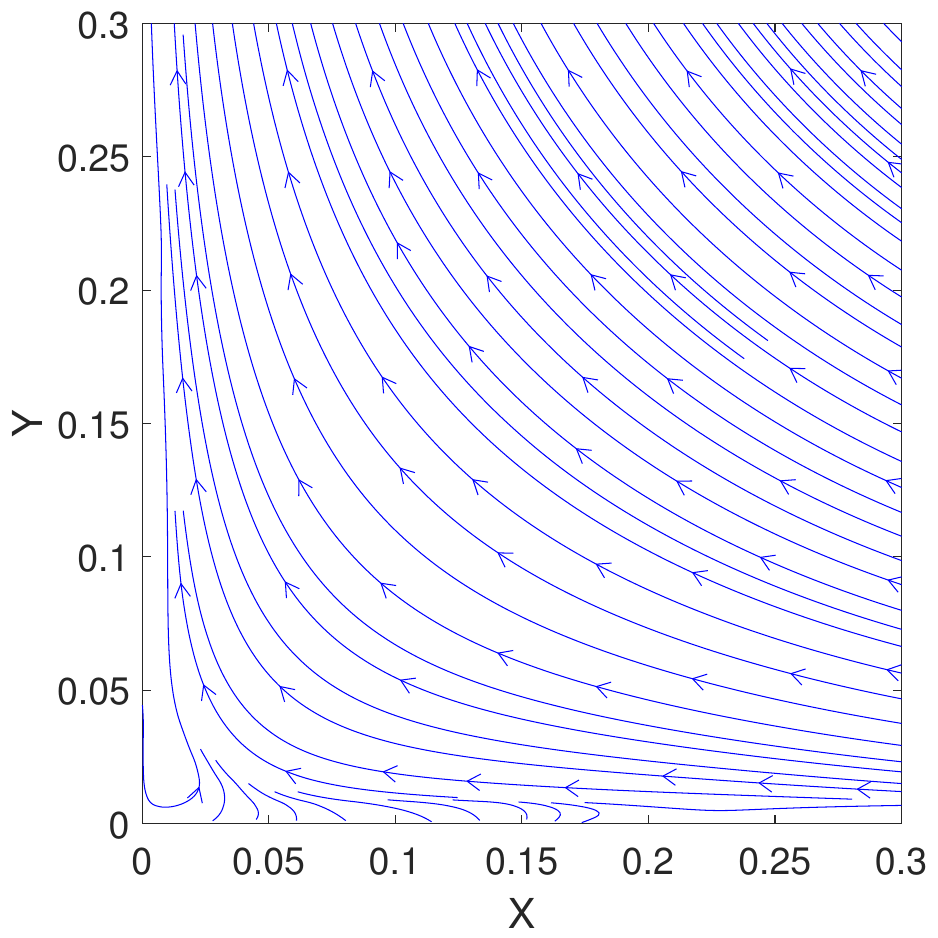} &
    \includegraphics[scale=.23]{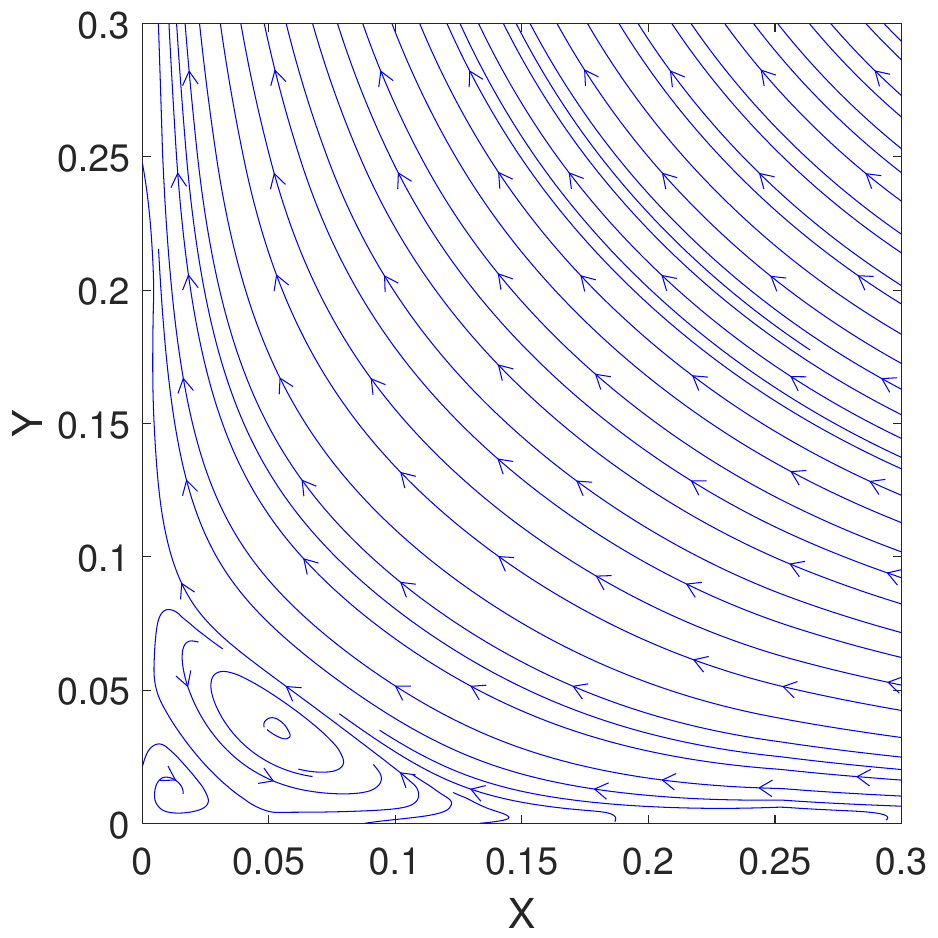} & \includegraphics[scale=.23]{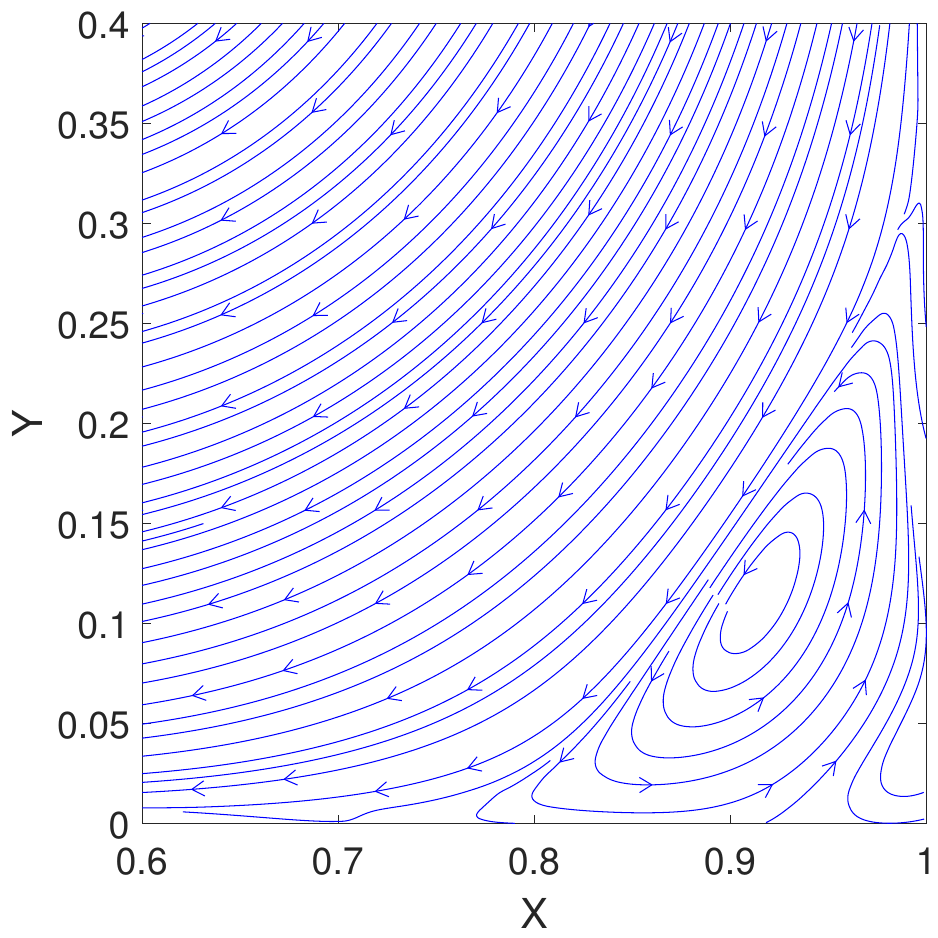} & \includegraphics[scale=.23]{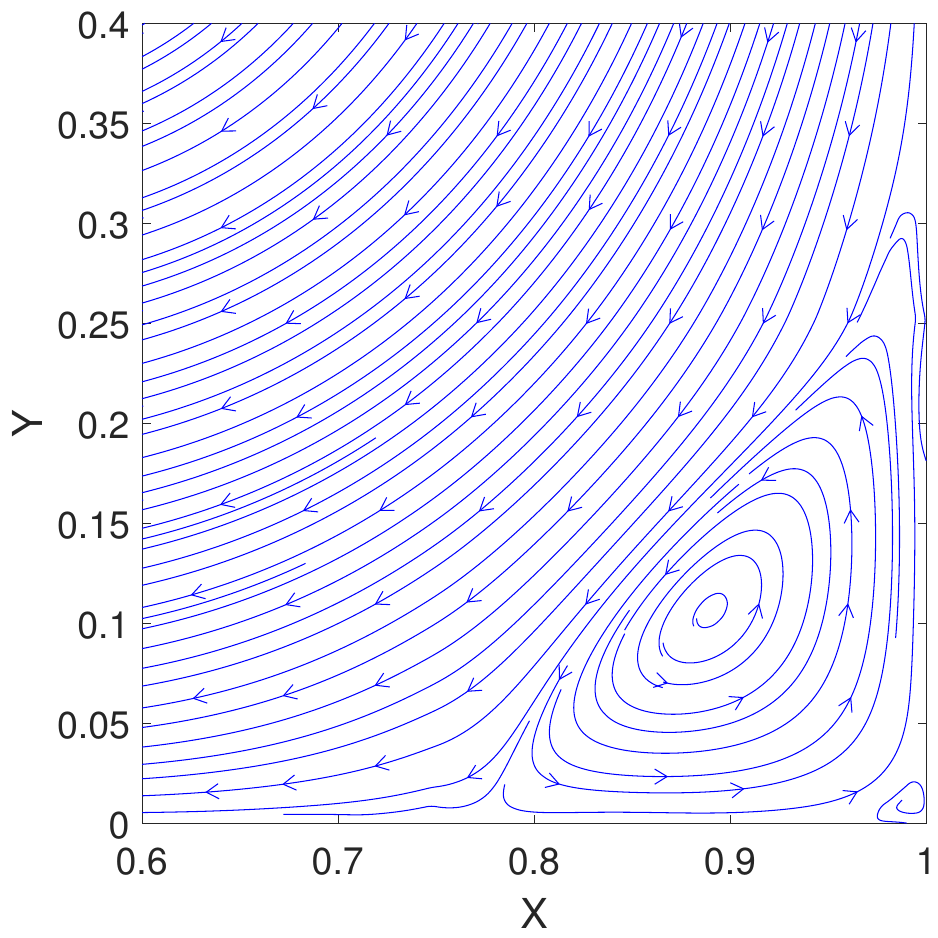} \\
    \textbf{(a)}  & \textbf{(b)} &  \textbf{(c)} & \textbf{(d)} \\
  \end{tabular}
  \caption{a.) Fluid streamlines using the standard FD--PINNs in the lower left corner of the cavity  with $120\times 120$ grid points. b.) Fluid streamlines using the present FD--PINNs in the lower left corner of the cavity with $50\times50$ sub--domain grid points. c.) Fluid streamlines using the standard FD--PINNs in the lower right corner of the cavity with with $120\times 120$ grid points. d.) Fluid streamlines using the present FD--PINNs in the lower right corner of the cavity with $50\times50$ sub--domain grid points.}
  \label{fig:8}
\end{figure}

   \begin{table}[H]\label{tab:6}
   \centering
   \caption{\label{tab:6}The MSE$_v$ comparison between the present and the standard FD--PINNs for the velocity $v$ at several horizontal lines on $[0,1]\times[0.91,.99]$ of the cavity for the Reynolds number $Re=1000$.}
   \vspace*{3mm}
\renewcommand\arraystretch{1.1}
\resizebox{\textwidth}{!}{
\begin{tabular}{ccccc}
\hline

$y$       &  The present FD--PINNs       &  The present FD--PINNs & The present FD--PINNs                        &  The standard FD--PINNs     \\                                                                           
&($25\times25$ sub--domain grid points )  & ($50\times50$ sub--domain grid points ) & ($100\times100$ sub--domain grid points ) & ($100\times100$ grid points)
                                                                                \\ \hline
    \\ \hline
    0.91      &  6.0343e-05   & 2.8282e-05      &  4.1387e-05 &      7.0913e-05   \\
      
        0.92    &   7.3488e-05  & 2.0961e-05      &    3.3037e-05 &       8.7717e-05    \\
     
       0.93    &     9.5251e-05  &  1.6014e-05       &      2.5434e-05 &    1.1124e-04
     \\
     
       0.94     &   1.2577e-04  &  1.4475e-05     &      1.8597e-05 &      1.4240e-04
  \\
       
       0.95  &    1.6351e-04   & 1.8090e-05      &      1.2792e-05 &        1.8296e-04  \\
      
       0.96    &  2.0897e-04   & 2.8808e-05      &      8.5852e-06 &        2.4448e-04
   \\
        
          0.97  &  3.0069e-04  &  4.5176e-05     &     6.6566e-06 &      3.9274e-04   \\
       
          0.98    &  6.8093e-04 &  6.6661e-05      &           7.7844e-06 &      9.0327e-04   \\
         
          0.99  &      2.2000e-03    &  3.6345e-04   &         1.2136e-05
 &       2.6000e-03   \\
\hline
\end{tabular}}

\end{table}
To show the effectiveness of the present technique, we consider the case where the secondary vortex completely disappeared from the lower left corner of the cavity for the standard FD--PINNs and for which consider the lid--driven cavity of size $[0,~1]\times[0,~1]$ and divided into $120\times120$ grid points. We consider 16 sub--domains, and each of them is divided into $50\times50$ grid points; with that consideration, by implementing the present technique, we successfully generate the secondary vortex in the lower left corner; see Figure \ref{fig:8}(b). Whereas, although the secondary vortex in the lower right corner appears for FD--PINNs with $120\times120$ grid points, but it is significantly less accurate than the present technique; see Figures \ref{fig:8}(c,d).

\subsection{Simulation of lid--driven cavity flow via the present FD--PINNs for Re=400}
\begin{figure}[H]
\centering
   \begin{tabular}{@{}cc@{}}
    \includegraphics[scale=.3]{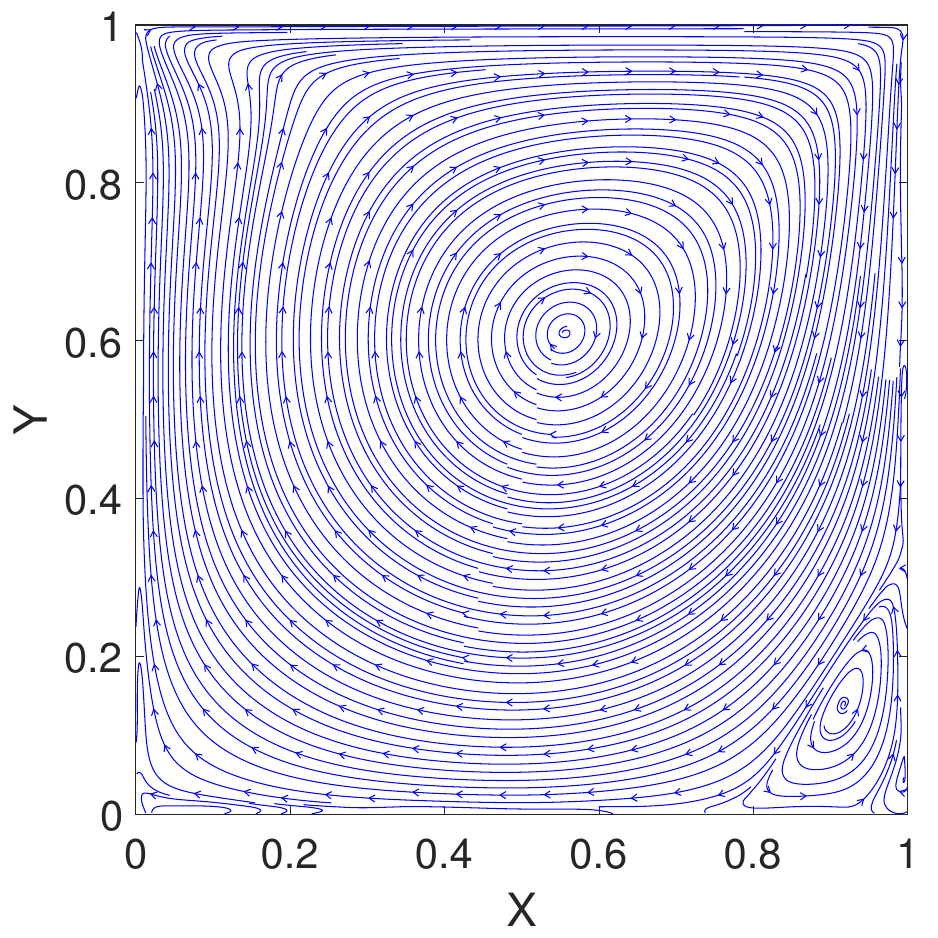} &
    \includegraphics[scale=.3]{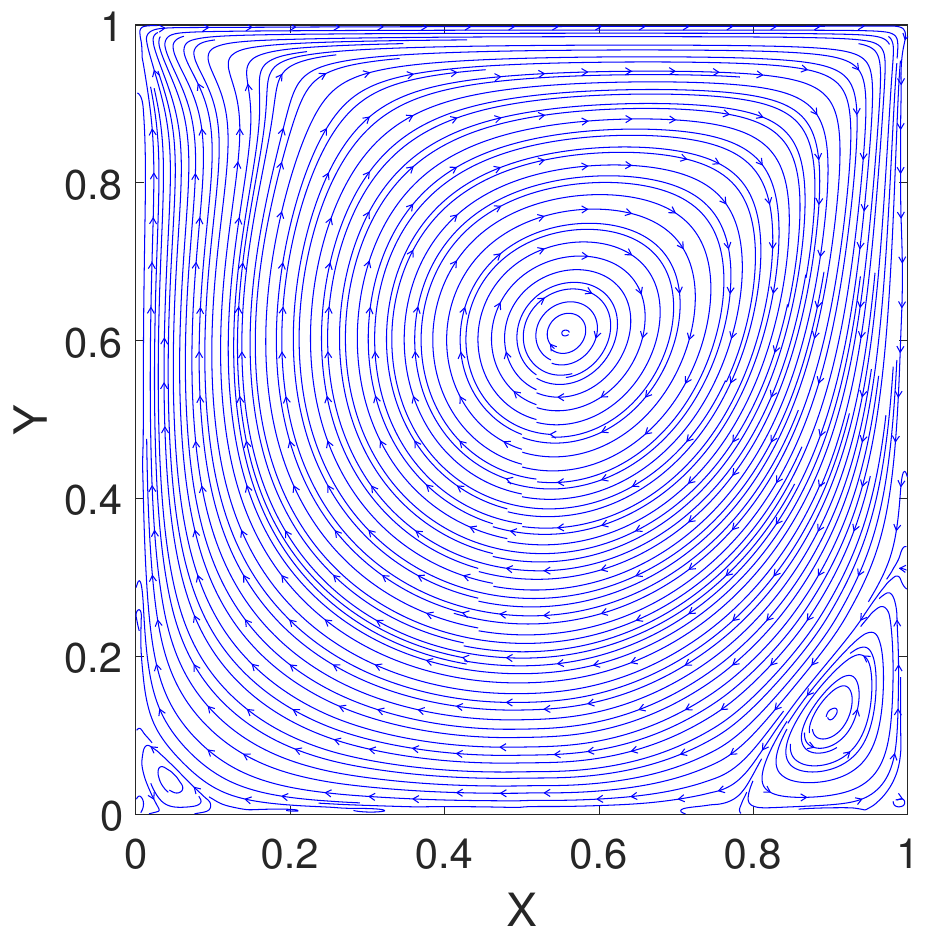} \\
    \textbf{(a)}  & \textbf{(b)} \\
  \end{tabular}
  \caption{a.) Fluid streamlines for the Reynolds number Re=400 using the standard FD--PINNs  with $100\times 100$ grid points. b.) Fluid streamlines for the Reynolds number Re=400 using the present FD--PINNs with $25\times25$ sub--domain grid points.}
  \label{fig:7}
\end{figure}

In this section, the impact of the present FD--PINNs is shown for the case of low Reynolds number flow, wherein Re=400 is considered for the simulation. Each sub--domain is discretized with $25\times 25$ grid points. A pictorial representation of the fluid flow for the Reynolds number Re=400 is shown in Figure \ref{fig:7}. It can be seen that the vortex in the lower left corner of the cavity completely vanishes for the standard FD--PINNs with $100\times 100$ grid points, whereas the present technique generates the vortex very efficaciously. The accuracy also improved significantly in the right bottom corner of the cavity by applying the present technique with only $25\times25$ sub--domain grid points. By applying a very small number of sub--domain grid points, the solution accuracy is improved drastically near the walls and in the corners of the lid--driven cavity, and are formed accurate secondary vortices in both corners. In contrast, inside the lid--driven cavity, the solution accuracy decreases for the small amount of sub--domain grid points (e.g., $25 \times 25$ grid points) like Re=1000. Therefore, to improve the solution accuracy only near the walls and generate secondary vortices in the corners of the cavity, the present FD--PINNs is sufficient to apply only in the regions near boundaries with a small number of sub--domain grid points.

\begin{figure}[H]
\centering
 \begin{tabular}{@{}c@{}}
    \includegraphics[width=.99\textwidth]{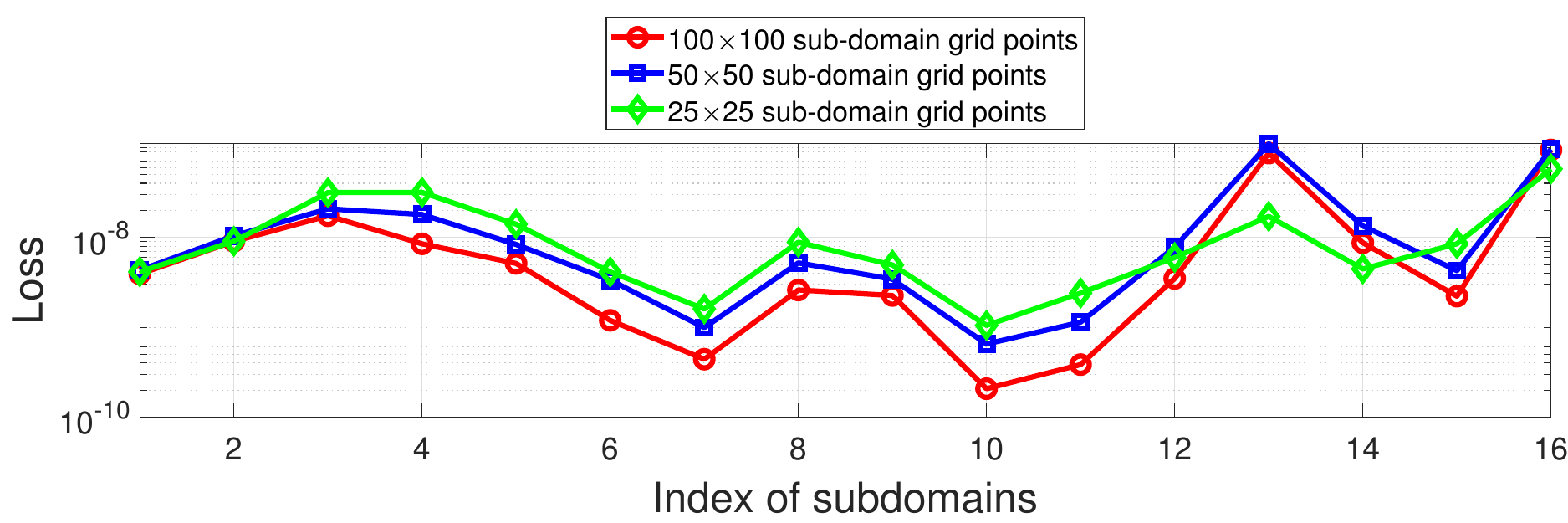}\\
 %   \textbf{(a)}\\
%    \includegraphics[width=.55\textwidth]{Loss_50.pdf} \\
%    \textbf{(b)}\\
%    \includegraphics[width=.55\textwidth]{Loss_25.pdf} \\
%    \textbf{(c)}\\
    \end{tabular}
  \caption{Final loss of the present FD--PINNs with $25\times25$, $50\times 50$ and $100\times 100$ sub--domain grid points.}
  \label{fig:9}
\end{figure}

  Figure \ref{fig:9} reflects the final normalized loss of the present FD--PINNs for all sub--domain with $100\times100$, $50\times50$, and $25\times25$ sub--domain grid points, respectively. It can be seen that the final normalized loss of all numbers of sub--domain grid points is significantly smaller ($<10^{-7}$), indicating that the networks for each sub--domain satisfy the governing equations very effectively. However, it is worth emphasizing that a lower value of the loss function does not always ensure a better solution than a slightly higher loss function; the accuracy of the solution is dependent on several other phenomena as well, which is also confirmed in \cite{jiang2023}. 

\section{Conclusions}\label{sec:4}

To mitigate the complexity of generating accurate secondary vortices in the lower corners of the cavity and improve solution accuracy in the boundary region of the cavity, an approach based on DDM and FD--PINNs is presented without taking into account the known solutions obtained from classical numerical methods. Instead of training new networks parallelly, networks are trained for each sub--domain completely independently, without the necessity of interphase conditions between two sub--domains. Boundary conditions for each sub--domain are generated from the solution of the entire domain. There are other ways to improve the accuracy of the solution and generate accurate secondary vortices in the lower corners of the lid--driven cavity. One advantage of this technique is that it is quite simple and easy to handle the boundary conditions of the sub--domain. The disadvantage of the technique is that the solution of the sub--domain depends on the solution of the entire domain, i.e., the solution accuracy of each sub--domain decreases if the solution of the entire domain is not sufficiently accurate. Finally, we emphasizing that this approach is straightforward in finding accurate solutions near the boundary of the cavity and generating secondary vortices in the lid--driven cavity by solving the Navier--Stokes equations.  Future directions are: first, coupling of DDM with FD--PINNs for solving Navier--Stokes equations in complex geometry and for higher ranges of Reynolds numbers. Furthermore, without interphase conditions, the implementation and computational complexity may be reduced for 3D problems. Therefore, secondly, testing the accuracy of the solution for 3D geometrical structures is of interest in future studies.

\bibliographystyle{ieeetr}
\bibliography{PINNs_FD.bib}

\end{document}